\catcode`\@=11

\newif\if@fewtab\@fewtabtrue

{\count255=\time\divide\count255 by 60
\xdef\hourmin{\number\count255}
\multiply\count255 by-60\advance\count255 by\time
\xdef\hourmin{\hourmin:\ifnum\count255<10 0\fi\the\count255}}
\def\ps@draft{\let\@mkboth\@gobbletwo
    \def\@oddhead{}
    \def\@oddfoot
       {\hbox to 7 cm{$\scriptstyle Draft\ version:\ \draftdate$
       \hfil}\hskip -7cm\hfil\rm\thepage \hfil}
    \def\@evenhead{}\let\@evenfoot\@oddfoot}

\def\ceqno{\global\@fewtabfalse
    \ifcase\@eqcnt \def\@tempa{& & &}\or \def\@tempa{& &}
      \or \def\@tempa{&}
      \or\def\@tempa{}\fi\@tempa
{\rm(\theequation)}}

\def\aeqno#1{\global\@fewtabfalse
    \ifcase\@eqcnt \def\@tempa{& & &}\or \def\@tempa{& &}
      \or \def\@tempa{&}
      \or\def\@tempa{}\fi\@tempa
{\rm(\theequation,#1)}}

\def\label#1{\ifnum\draftcontrol=1
 \global\def\draftnote{$\scriptstyle #1$}\fi
 \@bsphack\if@filesw {\let\thepage\relax
   \def\protect{\noexpand\noexpand\noexpand}%
\xdef\@gtempa{\write\@auxout{\string
      \newlabel{#1}{{\@currentlabel}{\thepage}}}}}\@gtempa
   \if@nobreak \ifvmode\nobreak\fi\fi\fi
  \@esphack}

\def\alabel#1#2{\label{#1}\global\@fewtabfalse
    \ifcase\@eqcnt \def\@tempa{& & &}\or \def\@tempa{& &}
      \or \def\@tempa{&}
      \or\def\@tempa{}\fi\@tempa
{\hbox to 3cm{\phantom{\rm(\theequation,#2)}
\draftnote \hfil}\hskip -3cm {\rm(\theequation,#2)}}}

\def\clabel#1{\label{#1}\global\@fewtabfalse
    \ifcase\@eqcnt \def\@tempa{& & &}\or \def\@tempa{& &}
      \or \def\@tempa{&}
      \or\def\@tempa{}\fi\@tempa
{\hbox to 3cm{\phantom{\rm(\theequation)}
\draftnote \hfil}\hskip -3cm{\rm(\theequation)}}}

\def\eqnarray{\def\draftnote{{}}\global\@fewtabtrue
\stepcounter{equation}\let\@currentlabel=\theequation
\global\@eqnswtrue
\global\@eqcnt\z@\tabskip\@centering\let\\=\@eqncr
$$\halign to \displaywidth\bgroup\@eqnsel\hskip\@centering\@eqcnt\z@
  $\displaystyle\tabskip\z@{##}$&\global\@eqcnt\@ne
  \hskip 1\arraycolsep \hfil${##}$\hfil
  &\global\@eqcnt\tw@ \hskip 1\arraycolsep
$\displaystyle\tabskip\z@{##}$
\hfil  \tabskip\@centering&\global\@eqcnt\thr@@\llap{##}\tabskip\z@
\cr}

\def\endeqnarray{\@@eqncr\egroup
      \global\advance\c@equation\m@ne$$\global\@ignoretrue}

\def\@eqnnum{\hbox to 3cm{\phantom{\rm(\theequation)} \draftnote
                         \hfil}\hskip -3cm {\rm(\theequation)}}

\def\@@eqncr{\let\@tempa\relax
    \ifcase\@eqcnt \def\@tempa{& & &}\or \def\@tempa{& &}
      \or \def\@tempa{&}
      \or\def\@tempa{}
\fi\@tempa
\if@eqnsw
\if@fewtab\@eqnnum\fi
\stepcounter{equation}\fi\global
\@eqnswtrue\global\@eqcnt\z@\global\@fewtabtrue\cr}

\def\draftcite#1{\ifnum\draftcontrol=1#1\else{}\fi}

\def\@lbibitem[#1]#2{\item{}\hskip -3cm \hbox to 2cm
{\hfil$\scriptstyle\draftcite{#2}$}\hskip
1cm[\@biblabel{#1}]\if@filesw
     {\def\protect##1{\string ##1\space}\immediate
      \write\@auxout{\string\bibcite{#2}{#1}}}\fi\ignorespaces}

\def\@bibitem#1{\item\hskip -3cm \hbox to 2cm
{\hfil $\scriptstyle\draftcite{#1}$}\hskip 1cm
\if@filesw \immediate\write\@auxout
       {\string\bibcite{#1}{\the\value{\@listctr}}}\fi\ignorespaces}

\def\nsection#1{\section{#1}\setcounter{equation}{0}}

\font\tendl=msbm10  scaled \magstep1
\font\sevendl=msbm7 scaled \magstep1
\font\fivedl=msbm5 scaled \magstep1
\font\tengl=eufm10  scaled \magstep1
\font\sevengl=eufm7 scaled \magstep1
\font\fivegl=eufm5 scaled \magstep1

\newfam\dlfam  
\textfont\dlfam=\tendl \scriptfont\dlfam=\sevendl
\scriptscriptfont\dlfam=\fivedl
\newfam\glfam  
\textfont\glfam=\tengl \scriptfont\glfam=\sevengl
\scriptscriptfont\glfam=\fivegl

\def\draftdate{\number\month/\number\day/\number\year\ \ \ \hourmin }

\global\def\draftcontrol{0}
\catcode`\@=12
\def\tilde{\widetilde}
\def\hat{\widehat}
\documentclass[aps,amsfonts,amsmath,amssymb,floatfix,nofootinbib]{revtex4}
\usepackage{epsfig}
\usepackage{float}
\usepackage{rotating}
\usepackage[centerlast]{subfigure}
\usepackage{bm}
\usepackage{amssymb}
\usepackage{comment}

\renewcommand{\theequation}{\arabic{section}.\arabic{equation}}
\newcommand{\ii}{\mathrm{i}}
\newcommand{\be}{\begin{eqnarray}}
\newcommand{\en}{\end{eqnarray}\vs 0.5 cm}

\newcommand{\Id}{{I\hspace{-0.04cm}d}}

\newcommand{\vs}{\vskip}

\newcommand{\Zb}{{\mathbb Z_2}}

\newcommand{\qq}{\begin{eqnarray}}

\newcommand{\ee}{{\rm e}}

\newcommand{\qqq}{\end{eqnarray}}

\newcommand{\tr}{\hbox{tr}}

\newcommand{\Hol}{{H\hspace{-0.02cm}ol}}

\newcommand{\CD}{{\cal D}}

\newcommand{\CF}{{\cal F}}
\newcommand{\CG}{{\cal G}}

\newcommand{\CK}{{\cal K}}
\newcommand{\CL}{{\cal L}}

\newcommand{\CN}{{\cal N}}

\newcommand{\CP}{{\cal P}}
\newcommand{\CQ}{{\cal Q}}
\newcommand{\CR}{{\cal R}}

\newcommand{\CT}{{\cal T}}

\newcommand{\CZ}{{\cal Z}}

\newcommand{\m}{\hspace{0.025cm}}

\newcommand{\WZ}{{W\hspace{-0.05cm}Z}}

\newcommand{\KM}{{K\hspace{-0.05cm}M}}
\begin{document}
\title{{\Large\bf{Square root of gerbe holonomy and invariants of
time-reversal-symmetric topological insulators}}}
\author{Krzysztof Gaw\c{e}dzki\footnote{directeur de recherche \'em\'erite, 
\,email: kgawedzk@ens-lyon.fr}}
\affiliation{Universit\'e de Lyon, ENS de Lyon, Universit\'e Claude Bernard, 
CNRS\\ Laboratoire de Physique, F-69342 Lyon, France\\
}


\maketitle

\vskip -0.2cm

\centerline{\bf\small ABSTRACT}
\vskip 0.2cm

\noindent The Feynman amplitudes with the two-dimensional Wess-Zumino
action functional have a geometric interpretation as bundle gerbe holonomy.
We present details of the construction of a distinguished
square root of such holonomy and of a related $\,3d$-index and
briefly recall the application of those to the building of topological
invariants for time-reversal-symmetric two- and three-dimensional crystals,
both static and periodically forced.

\

\vskip 0.3cm

\nsection{Introduction}
\label{sec:intro}

\noindent The central theme of the present paper is a specific
refinement of the two-dimensional Wess-Zumino (WZ) field-theoretic
action functional that finds its application in the construction
of invariants of the times-reversal-symmetric topological insulators
in two and three dimensions. The WZ functional has
appeared in the context of field theory anomalies \cite{WZ}.
A two-dimensional WZ action related to the chiral anomaly was used
in \cite{WittenNA} as an important component in
the construction of a particular conformal field theory, the so called
Wess-Zumino-Witten sigma model. As opposed to local action functionals,
the $2d$ WZ action $\,S_\WZ(\phi)\,$ of a classical field $\,\phi\,$
was originally defined by a local functional of its three-dimensional
extension. This resulted in $\,S_\WZ(\phi)\,$ defined only modulo
$\,2\pi$, \,leading, however, to the univalued Feynman
amplitude $\,\ee^{\ii S_\WZ(\phi)}\,$ (we set the Planck constant
$\,\hbar\,$ to $\,1\,$ here). The first attempts
to write local formulae for $\,S_\WZ(\phi)\,$ were based on cohomological
approaches [\onlinecite{Alvarez},\,\onlinecite{KG}]. In particular,
\cite{KG} realized that the Deligne cohomology \cite{Deligne} provided
the proper tool for such problems, permitting to define the WZ action
in more general situations. With the advent of the theory of bundle gerbes
[\onlinecite{Murray},\,\onlinecite{MS}], this approach has gained a geometric
interpretation:
the Feynman amplitudes $\,\ee^{\ii S_\WZ(\phi)}\,$ got the interpretation
of the ``bundle gerbe holonomy'' [\onlinecite{CMM},\,\onlinecite{GR}].
\vskip 0.1cm

The refinement of the WZ action that we shall discuss here will permit
to fix a square root of the WZ amplitude for ``equivariant'' fields
$\,\phi\,$ that intertwine an orientation preserving
involution of the closed surface on which they are defined with an
involution in the target space. Giving a unique value to the square root
of the WZ amplitude of equivariant fields will require to define their
WZ action modulo $\,4\pi\,$ rather than only modulo $\,2\pi$.
\,In a somewhat implicit way, the square root of the WZ amplitude was used 
in [\onlinecite{CDFG},\,\onlinecite{CDFGT}] to define dynamical torsion
invariants of periodically forced crystalline systems with time-reversal
symmetry. In the case of such Floquet systems, the surface was the
Brillouin 2-torus equipped with the involution reversing the sign
of the quasimomentum and the target space was the unitary group
with the involution corresponding to the time reversal. On the way, it
was shown in those references that the Fu-Kane-Mele invariant
[\onlinecite{KaneMele},\,\onlinecite{FK}] of the static time-reversal-symmetric
topological insulators may be expressed as the square root of a WZ
amplitude.
\vskip 0.1cm

From the point of view of gerbes, \,in order to fix the
square root of the bundle gerbe holonomy describing the WZ
amplitude of equivariant fields, one needs an additional structure
expressing the equivariance of the gerbe under the target-space involution.
The situation bears some similarity to that where the WZ amplitudes
are extended to fields
defined on non-oriented surfaces, studied previously in
[\onlinecite{SSW},\,\onlinecite{GSW},\,\onlinecite{NS}]. There
are, however, some important differences. The most
notable of those is that, in general, the orientation preserving involutions
of surfaces possess fixed points that require special treatment,
unlike the orientation-reversing involutions of oriented covers of
non-oriented surfaces. The aspects of the bundle gerbe theory needed to
define the square root of gerbe holonomy
were reviewed in much detail in the lecture notes \cite{GenWar} of
the present author, together with their applications to two- and
three-dimensions topological insulators and Floquet systems. What was
omitted there, however, was the proofs that the presented formulae for
the square root of the gerbe holonomy and for a related $\,3d$-index
define those quantities in an unambiguous way. The main goal of the present
paper is to fill that gap. We use here a different but equivalent
presentation of the additional structure on the gerbe needed for the
construction of the square root of gerbe holonomy. This permits
to streamline the proofs.
\vskip 0.1cm

The paper is organized as follows. In Sec.\,\ref{sec:bdl_gerbes},
we recall the definition of bundle gerbes and in Sec.\,\ref{sec:gerbe_hol},
that of gerbe holonomy. Sec.\,\ref{sec:equiv_gerbes} introduces the
notion of an equivariant extension of a gerbe with respect to
a $\,\Zb$-action defined by an involution on the base space.
The central Sec.\,\ref{sec:sqrt_hol} presents a local formula for
the square root of gerbe holonomy and proves that it
determines the latter in an unambiguous way under some topological
conditions. In Sec.\,\ref{sec:homot_form},
we establish a non-local formula for the same quantity. 
Such a formula was employed in \cite{CDFGT} as the definition
of the square root of the WZ amplitude. Sec.\,\ref{sec:loc_data}
describes the equivariant extension of gerbes and the square
root of gerbe holonomy in terms of local data. Sec.\,\ref{sec:3d_index}
presents a formula, involving the square root of gerbe holonomy, for the
$\,3d$-index and proves that the latter is unambiguously defined. 
The last two sections cover the subjects discussed in detail
in \cite{GenWar} and are included for completeness.
In Sec.\,\ref{sec:basic_gerbe}, we briefly describe how the general
scheme can be extended to the case of the basic gerbe on the
unitary group with the involution given by the time reversal
and, in Sec.\,\ref{sec:appli_TI}, we summarize the application
of such an extension to the construction of invariants of
time-reversal-symmetric crystalline systems, both static and periodically
driven. Appendix discusses the relation between the equivariant extension
of gerbes used in the present paper and the equivariant structure
on gerbes employed in \cite{GenWar}, making explicit the relation
between the constructions of both papers.
\vskip 0.2cm

\noindent{\bf Acknowledgements}. \,I thank David Carpentier, Pierre Delplace,
Michel Fruchart and Cl\'ement Tauber for the collaboration on physical
aspects of the topics discussed in this paper. I am grateful to Pavol
$\check{\rm S}$evera for advocating the way of looking at equivariant
gerbes that was employed here. I have profited from a discussion with
Etienne Ghys about surfaces with involution. I also thank the
organizers of the IGA/AMSI 2016 Workshop in Adelaide for the
invitation that provided an opportunity to present results related
to this paper.

\nsection{Bundle gerbes}
\label{sec:bdl_gerbes}

\noindent Bundle gerbes are examples of higher structures, 1-degree higher
than line bundles. They were introduced by M. K. Murray \cite{Murray} in
1996, see also \cite{Murray-Stevenson}, as geometric examples of more 
abstract gerbes of J. Giraud \cite{Giraud} and J.-L. Brylinski 
\cite{Brylinski}.
Below, we shall only consider bundle gerbes and line bundles
equipped with hermitian structure and unitary connection without further
mention. The aspect of bundle gerbes that we shall be interested in here
is that they provide local formulae for topological Feynman amplitudes
of the Wess-Zumino (WZ) type for two-dimensional classical
fields, as already mentioned in Introduction.
\vskip 0.1cm

Let us start by recalling some notations. We shall work in the category
of smooth manifolds. If $\,\pi:Y\rightarrow M\,$ then by $\,Y^{[n]}\,$ 
we shall denote the subset
of $\,Y^{n}\,$ composed of $\,(y_i)_{i=1}^n\,$ with all $\,\pi(y_i)\,$ equal.
For a sequence $\,(i_1,\dots,i_m)\,$ with $\,1\leq i_j\leq n$, 
$\,p_{i_1\dots i_m}\,$ will denote the map from $\,Y^{[n]}\,$ to $\,Y^{[m]}\,$
such that $\,p_{i_1\dots i_m}(y_1,\dots,y_n)=(y_{i_1},\dots,y_{i_m})$.
\,If $\,\pi\,$ is a submersion of manifolds then $\,Y^{[n]}$ is 
a submanifold of $\,Y^n\,$ and the maps $\,p_{i_1\dots i_k}\,$ are smooth.
\vskip 0.3cm

\noindent{\bf Definition 1} \cite{Murray}{\bf.} \,A bundle gerbe $\,\CG\,$ 
(below, a ``gerbe'' for short) over $\,M\,$ is a 
quadruple $\,(Y,B,\CL,t)$, \,where $\,\pi:Y\rightarrow M\,$ is a surjective 
submersion, $\,B\,$ is a real 2-form
on $\,Y\,$ (called the curving), $\,\CL\,$ is a line bundle over 
$\,Y^{[2]}\,$ with curvature 2-form $\,F_\CL=p_2^*B-p_1^*B$, \,and $\,t\,$ is 
a line-bundle isomorphism over $\,Y^{[3]}\,$
\qq
t:p_{12}^*\CL\otimes p_{23}^*\CL\longrightarrow p_{13}^*\CL\,,
\label{t}
\qqq
acting fiber-wise\footnote{We denote by $\,\CL_{y_1,y_2}\,$ the fiber
of $\,\CL\,$ over $\,(y_1,y_2)\in Y^{[2]}$.} as $\,
\CL_{y_1,y_2}\otimes\CL_{y_2,y_3}\mathop{\longrightarrow}\limits^t
\CL_{y_1,y_3}\,$ for $\,(y_1,y_2,y_3)\in Y^{[3]}$, \,that defines 
an (associative) groupoid multiplication on 
$\,\CL\,\substack{\rightarrow\\\rightarrow}\,Y$.
\vskip 0.3cm

The condition on the curving 2-form implies that $\,p_1^*dB=p_3^*dB\,$ so 
that $\,dB=\pi^*H\,$ for some closed 3-form $\,H\,$ on $\,M\,$ called the 
curvature of the gerbe $\,\CG$. \,The isomorphism $\,t\,$ provides 
a canonical trivialization of the line bundle $\,d^*\CL$, where $\,d\,$ 
is the diagonal embedding of $\,Y\,$ into $\,Y^{[2]}\,$ and a canonical 
isomorphism of $\,\sigma^*\CL\,$ with $\,\CL^{-1}$, \,where 
$\,\sigma(y_1,y_2)=(y_2,y_1)\,$ and $\,\CL^{-1}\,$ denotes the line bundle 
dual to $\,\CL$. \,In particular, $\,\CL_{y,y}\cong\mathbb C\,$ and 
$\,\CL_{y_1,y_2}^{-1}\cong\CL_{y_2,y_1}\,$ canonically.

\nsection{Bundle gerbe holonomy}
\label{sec:gerbe_hol}

\noindent Let $\,\Sigma\,$ be a closed oriented surface. If 
$\,\CG=(Y,B,\CL,t)\,$ is a gerbe over $\,M\,$ and 
$\,\phi:\Sigma\rightarrow M\,$ then one may associate to $\,\phi\,$
a phase in $\,U(1)\,$ denoted $\,Hol_\CG(\phi)\,$ and called the holonomy 
of $\,\CG\,$ along $\,\phi$. \,We shall need an explicit representation
of such a phase. 
\vskip 0.1cm

\begin{figure}[!h]
\vskip -0.3cm
\begin{center}
\leavevmode
 \includegraphics[width=9.7cm,height=4.3cm]{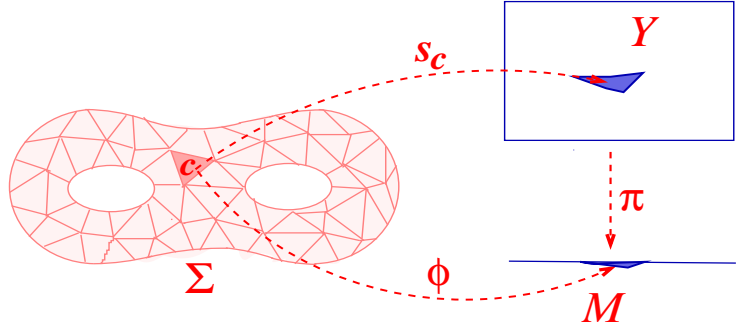}
\vskip -0.1cm
\caption{Triangulation of $\,\Sigma\,$ for gerbe holonomy calculation}
\label{fig:triang0}
\end{center}
\vskip -0.2cm
\end{figure}

To this end, let us choose a triangulation of $\,\Sigma\,$ composed 
of triangles $\,c\,$ (with orientation inherited from $\,\Sigma$), 
\,edges $\,b\,$ and vertices $\,v$, \,see Fig.\,\ref{fig:triang0}. 
\,We suppose that it is sufficiently fine so that one may choose maps 
$\,s_c:c\rightarrow Y\,$ and $\,s_b:b\rightarrow Y\,$ and elements
$\,s_v\in Y\,$ such that  
\qq
\pi\circ s_c=\phi|_c\,,\qquad \pi\circ s_b=\phi|_b\,,\qquad\pi(s_v)=\phi(v)\,.
\label{lsc}
\qqq

\begin{figure}[!h]
\vskip 0.2cm
\begin{center}
\leavevmode
 \includegraphics[width=3.8cm,height=3cm]{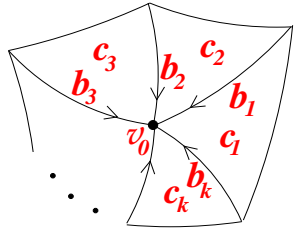}
\vskip -0.05cm
\caption{Triangulation of $\,\Sigma\,$ around vertex $\,v_0$}
\label{fig:aroundv0}
\end{center}
\vskip -0.5cm
\end{figure}

\noindent Then the holonomy of $\,\CG\,$ along $\,\phi\,$ may be given 
by the expression \cite{GR}
\qq
Hol_\CG(\phi)\,=\,\ee^{\ii\sum\limits_c\int_cs_c^*B}\hspace{-0.1cm}
\mathop{\otimes}\limits_{b\subset c}hol_\CL(s_c|_b,s_b)\,,
\label{HolCG}
\qqq
where we use a slightly abusive notation in which $\,hol_\CL({\ell})\,$
stands for the parallel transport in the line bundle $\,\CL\,$ along
the curve $\,{\ell}\,$ in $\,Y^{[2]}$, \,a linear map from the fiber 
of $\,\CL\,$ over the initial point of $\,\ell\,$ to the one over the final 
point. {\it A priori}, the expression on the right hand side of
(\ref{HolCG}) is an element of the line
\qq
\mathop{\otimes}\limits_{v\in b\subset c}\CL_{s_c(v),s_b(v)}^{\pm1}\,,
\label{line}
\qqq
where the minus power (the dual line) is chosen if $\,v\,$ has a negative
orientation, i.e. is the initial point of the edge $\,b\,$ with orientation 
inherited from $\,c$. \,The groupoid structure on $\,\CL\,$ defined by the 
isomorphism $\,t\,$ of (\ref{t}), however, makes the line (\ref{line}) 
canonically isomorphic to $\,\mathbb C$. \,Indeed, for a fixed vertex 
$\,v_0\,$ as in Fig.\,\ref{fig:aroundv0}, 

\noindent we have
\qq
\mathop{\otimes}\limits_{v_0\in b\subset c}\CL_{s_c(v_0),s_b(v_0)}^{\pm1}
&\cong&\,\CL_{s_{c_1}(v_0),s_{b_1}(v_0)}\otimes\CL_{s_{b_1}(v_0),s_{c_2}(v_0)}
\otimes\CL_{s_{c_2}(v_0),s_{b_2}(v_0)}\otimes\CL_{s_{b_2}(v_0),s_{c_3}(v_0)}\cr
&&\hspace{-1cm}\otimes\,\CL_{s_{c_3}(v_0),s_{b_3}(v_0)}\otimes\ \cdots\cdots\ \otimes
 \CL_{s_{c_k}(v_0),s_{b_k}(v_0)}\otimes\CL_{s_{b_k}(v_0),s_{c_1}(v_0)}
\ \cong\ \CL_{s_{c_1(v_0)},s_{c_1(v_0)}}\ \cong\ \mathbb C\qquad
\label{6}
\qqq
and a cyclic permutation of terms does not change 
the isomorphism with $\,\mathbb C\,$
because of the associativity of the groupoid multiplication in $\,\CL$. 
\,Hence, the right hand side of (\ref{HolCG}) may be canonically viewed
as a complex number that, in fact, is a phase in $\,U(1)$.
\vskip 0.4cm

\noindent{\bf Proposition 1} \cite{GR}{\bf.} \ The phase associated to
the right hand side of (\ref{HolCG}) is independent of the choice of the
maps $\,s_c\,$ and $\,s_b\,$ and of the triangulation of $\,\Sigma$.
\vskip 0.3cm

\noindent{\bf Proof\,}\footnote{Such proofs may be also done 
in the cohomological language using local data for gerbes
that we discuss in Sec.\,\ref{sec:loc_data}, see
[\onlinecite{KG},\,\onlinecite{GR}].}{\bf.} 
\,We give here a brief proof of Proposition 1 since below we shall need
its refinements.
\vskip 0.1cm

\,1.\,\ If we change 
the map $\,s_{c_0}\,$ to $\,s'_{c_0}\,$ for a triangle $\,c_0\,$ then
\qq
\ee^{\ii\int_{c_0} s'^*_{c_0}B}\,=\,\ee^{\ii\int_{c_0} s_{c_0}^*B}\,
\ee^{\ii\int_{c_0}(s'^*_{c_0}B-s^*_{c_0}B)}\,=\,\ee^{\ii\int_{c_0}s_{c_0}^*B}\,
\ee^{\ii\int_{c_0}(s_{c_0},s'_{c_0})^*F_{\CL}}\,\cong\,\ee^{\ii\int_{c_0}s_{c_0}^*B}
\hspace{-0.1cm}
\mathop{\otimes}\limits_{b\subset c_0}hol_\CL(s_{c_0}|_b,s'_{c_0}|_b)\,,\quad
\label{10}
\qqq
where the last tensor product that belongs to 
$\,\mathop{\otimes}_{v\in b\subset c_0}\CL_{s_{c_0}(v),s'_{c_0}(v)}^{\pm1}\hspace{-0.1cm}
\cong\mathbb{C}\,$
gives rise to the holonomy of $\,\CL\,$ along the loop
$\,(s_{c_0},s'_{c_0})|_{\partial c_0}\,$ in $\,Y^{[2]}$. \,Now the isomorphisms
$\,t\,$
map $\,hol_\CL(s_{c_0}|_b,s'_{c_0}|_b)\otimes hol_\CL(s'_{c_0}|_b,s_b)\,$ to
$\,hol_\CL(s_{c_0}|_b,s_b)\,$ so that one may identify
$\,\ee^{\ii\int_{c_0} s'^*_{c_0}B}\hspace{-0.1cm}\mathop{\otimes}_{b\subset c_0}
hol_\CL(s'_{c_0}|_b,s_b)\,$ with $\,\ee^{\ii\int_{c_0} s^*_{c_0}B}
\hspace{-0.1cm}\mathop{\otimes}_{b\subset c_0}hol_\CL(s_{c_0}|_b,s_b)\,$ and this 
identification commutes with the identification of both expressions
with numbers in $\,\mathbb C\,$ also based on applying isomorphisms
$\,t$, \,as the latter are associative.
\vskip 0.1cm

2.\ \,Similarly, if we change the map $\,s_{b_0}\,$ to $\,s'_{b_0}\,$
then we may identify $\,\mathop{\otimes}_{c\supset b_0}
hol_\CL(s_c|_{b_0},s'_{b_0})\,$ with $\,\mathop{\otimes}_{c\supset b_0}
(hol_\CL(s_c|_{b_0},s_{b_0})\otimes hol_\CL(s_{b_0},s'_{b_0}))\,$ using $\,t\,$ but
$\,\mathop{\otimes}_{c\supset b_0}hol_\CL(s_{b_0},s'_{b_0})\,$ is canonically
equal to 1 as the terms appear in dual pairs corresponding to two triangles
bordering the edge $\,b_0\,$ that induce on it opposite orientations.
\vskip 0.1cm

\begin{figure}[h]
\begin{center}
\vskip -0.1cm
\leavevmode
\hspace{1.1cm}
{%
      \begin{minipage}{0.43\textwidth}
        \includegraphics[width=3.7cm,height=2.7cm]{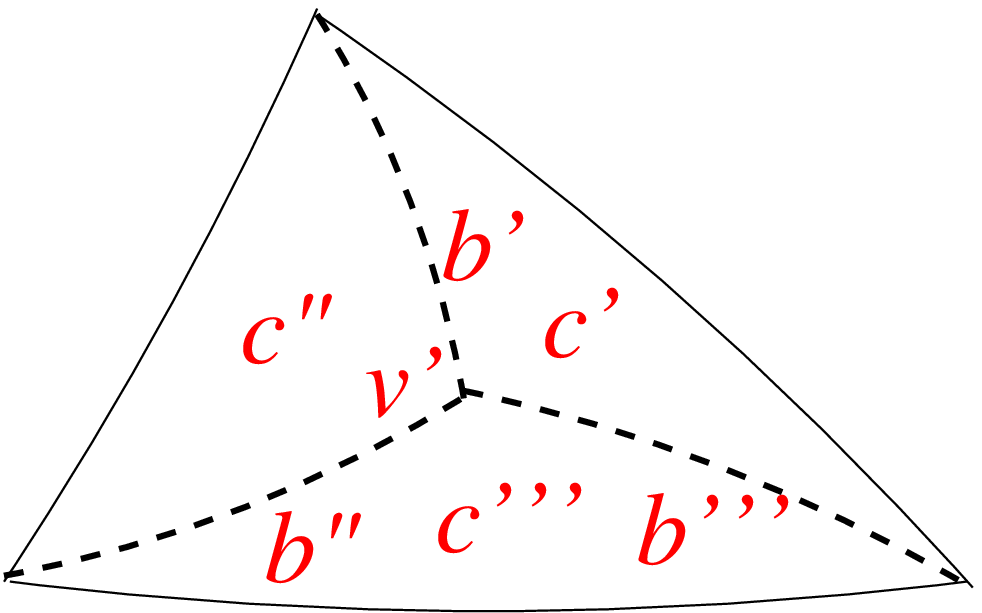}\\
        \vspace{-0.7cm} \strut
        \end{minipage}}
    \hspace*{-2cm}
{%
      \begin{minipage}{0.43\textwidth}
        \vspace{0.2cm}
        \includegraphics[width=3.7cm,height=2.7cm]{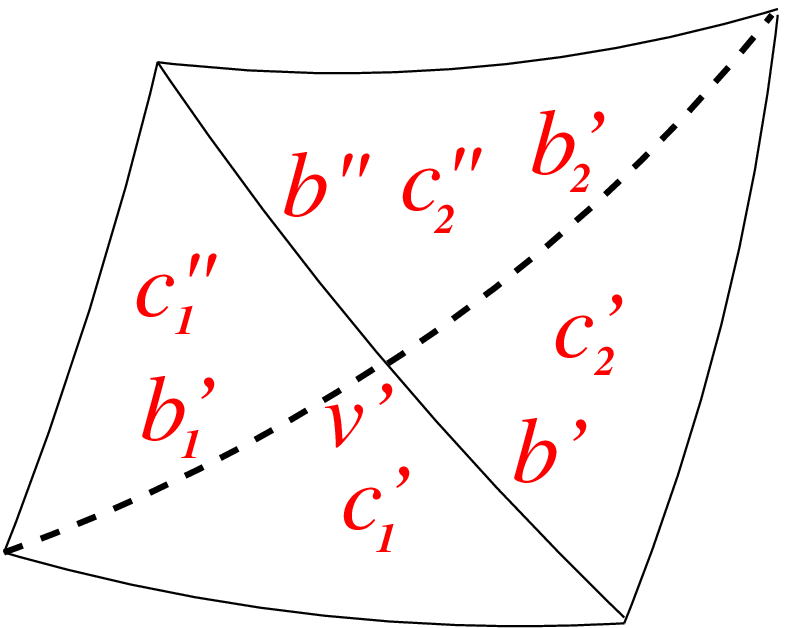}\\
      \vspace{-0.7cm} \strut
        \end{minipage}}
      \hspace*{15pt}
    \hspace{14pt}
\end{center}
\caption{Two ways of generating a finer triangulation of $\,\Sigma$}
\label{fig:subdiv}
\end{figure}

\noindent 3.\ \,To show that the phase associated to the right hand side
of (\ref{HolCG}) is independent of the triangulation, let us change the
latter by subdividing one of the triangles $\,c\,$ as on the left hand side
of \,Fig.\,\ref{fig:subdiv}, \,defining 
$\,s_{c'}=s_c|_{c'}$, $\,s_{b'}=s_c|_{b'}$, \,etc. Then the right hand side 
of (\ref{HolCG}) picks up additionally only trivial factors
$\,hol_\CL(s_c|_{b'},s_c|_{b'})\,$ etc. canonically identified with $\,1$.
\,Similarly, if we subdivide one of the edges $\,b\,$ and the neighboring
triangles $\,c_1,\,c_2\,$ as on the right hand side
of \,Fig.\,\ref{fig:subdiv}, \,choosing $\,s_{c'_1}=s_{c_1}|_{c'_1}$,
$\,s_{c''_1}=s_{c_1}|_{c''_1}$, $\,s_{b'_1}=s_{c_1}|_{b'_1}$,
$\,s_{c'_2}=s_{c_2}|_{c'_2}$, $\,s_{c''_2}=s_{c_2}|_{c''_2}$, 
$\,s_{b'_2}=s_{c_2}|_{b'_2}$ and $\,s_{b'}=s_b|_{b'}$, $\,s_{b''}=s_b|_{b''}\,$ then
the right hand side of (\ref{HolCG}) changes only by decomposing
$\,hol_\CL(s_{c_i}|_b,s_b)\,$ as $\,hol_\CL(s_{c_i}|_{b'},s_b|_{b'})
\otimes hol_\CL(s_{c_i}|_{b''},s_b|_{b''})\,$
and by adding factors canonically equal to $\,1\,$ and its numerical value
remains unchanged. The above shows that the phases associated to
the right hand side of (\ref{HolCG}) are equal for triangulations differing 
by two-dimensional Pachner moves \cite{Pachner} whose chains allow to 
relate any two triangulations of $\,\Sigma\,$ to a common third one.
\vskip -0.1cm

\hspace{14cm}$\blacksquare$

\vskip 0.1cm

\noindent{\bf Example 1.} \ If $\,\phi:\Sigma\mapsto M\,$ is a constant map
then $\,\Hol_\CG(\phi)=1$. \,Indeed, one may choose in this case all $\,s_c\,$
and $\,s_b\,$ to be constant and taking the same value and all contributions
to the right hand side of (\ref{HolCG}) become canonically equal to $\,1$.
\vskip 0.3cm  

If $\,\CD:\Sigma\rightarrow\Sigma\,$ a is diffeomorphism that preserves
or changes the orientation then, respectively,
\qq
\Hol_\CG(\phi)\,=\,\Hol_\CG(\phi\circ\CD)^{\pm1}.
\qqq
This follows by computing the right hand side with the triangulation
obtained from that used for the left hand side by application of
$\,\CD^{-1}\,$ and with the maps $\,s_c\circ\CD\,$ and $\,s_b\circ\CD$.
\vskip 0.4cm

\noindent{\bf Proposition 2.} \ If $\,\phi:\Sigma\rightarrow M\,$ has
an extension $\,\psi:\CT\rightarrow M\,$ to an oriented
compact $\,3$-manifold $\,\CT\,$ with boundary
$\,\partial\CT=\Sigma\,$ then for any gerbe $\,\CG\,$ with
curvature $\,H\,$ 
\qq
\Hol_\CG(\phi)\ =\ \ee^{\ii\int\limits_{\CT}\psi^*H}\,.
\label{SWZ}
\qqq
\vskip 0.3cm

\noindent{\bf Proof.} \ Let us triangulate
$\,\CT\,$ denoting by $\,h,c,b,v\,$ the corresponding simplices
assumed sufficiently small so that one may choose over them the lifts
$\,s_h$, $\,s_c$ and $\,s_b\,$ of $\,\psi\,$
to $\,Y$. \,The tetrahedra $\,h\,$ will be taken with the orientation
induced from $\,\CT$. \,Then
\qq
&&\ee^{\ii\int\limits_{\CT}\psi^*H}\,=\,
\ee^{\ii\sum\limits_{h}\int_hs_h^*dB}
\,=\,\ee^{\ii\sum\limits_{c\subset h}\int_cs^*_hB}\,
=\,\ee^{\ii\sum\limits_{c\subset h}\int_cs^*_cB}\,
\ee^{\ii\sum\limits_{c\subset h}\int_c(s^*_hB-s_c^*B)}\,=\,\ee^{\ii\sum\limits_{c\subset\Sigma}\int_cs^*_cB}\,
\ee^{\ii\sum\limits_{c\subset h}\int_c(s_c,s_h|_c)^*F_\CL}\cr\cr
&&=\,\ee^{\ii\sum\limits_{c\subset\Sigma}\int_cs^*_cB}\hspace{-0.05cm}
\prod\limits_{c\subset h}hol_\CL((s_c,s_h)|_{\partial c})
\ \cong\ \ee^{\ii\sum\limits_{c\subset\Sigma}\int_cs^*_cB}\hspace{-0.3cm}
\mathop{\otimes}\limits_{b\subset c
\subset h}\hspace{-0.15cm}
hol_\CL((s_c,s_h)|_b)\cr
&&\cong\ 
\ee^{\ii\sum\limits_{c\subset\Sigma}\int_cs^*_cB}\hspace{-0.3cm}
\mathop{\otimes}\limits_{b\subset c\subset h}\hspace{-0.1cm}
\Big(hol_\CL(s_c|_b,s_b)\otimes
hol_\CL(s_b,s_h|_b)\Big)
\,=\,\ee^{\ii\sum\limits_{c\subset\Sigma}\int_cs^*_cB}\hspace{-0.15cm}\mathop{\otimes}
\limits_{b\subset c\subset\Sigma}hol_\CL(s_c|_b,s_b)\,=\,\Hol_\CG(\phi)\,,
\label{Witeq}
\qqq
where the last but one equality arises since in the preceding expression
each term $\,hol_\CL(s_b,s_h|_b)\,$ appears twice with opposite
orientations of $\,b\,$ shared by two faces $\,c\,$ of $\,h\,$
and, similarly, each term $\,hol_\CL(s_c|_b,s_b)\,$ appears
twice with opposite orientations of $\,b\subset c\not\subset
\partial\CT=\Sigma\,$ corresponding to two $\,h\,$
sharing the face $\,c$.
\vskip 0.1cm

\hspace{13cm}$\blacksquare$
\vskip 0.3cm

\noindent{\bf Remark.} \,The homotopic formula (\ref{SWZ}) coincides with
Witten's definition of the WZ Feynman amplitude $\,\ee^{\ii S_\WZ(\phi)}\,$
[\onlinecite{WittenCA},\,\onlinecite{WittenNA}] which requires
that $\,\int_\CT\psi^*H\,$
for an extension $\,\psi\,$ of $\,\phi\,$ be well defined modulo $\,2\pi$.
\,This holds whenever $\,H\,$ is a curvature of a gerbe. There may, 
however, be $\,\phi\,$ with no extension $\,\psi\,$ and then 
$\,\Hol_\CG(\phi)\,$ cannot be defined this way. In such cases, which 
correspond to $\,M\,$ with non-trivial $2^{\rm nd}$ homology, 
$\,\Hol_\CG(\phi)\,$ depends also on the gerbe $\,\CG\,$ and not only 
on its curvature.

\nsection{Equivariance of gerbes under an involution}
\label{sec:equiv_gerbes}

\noindent Equivariant bundle gerbes were studied by several authors,
\,see [\onlinecite{Gomi},\,\onlinecite{Gomi1},\,\onlinecite{SSW},\,\onlinecite{GSW},\,\onlinecite{GSW1},\,\onlinecite{NS},\,\onlinecite{BenBassat},\,\onlinecite{MRSV}]. We shall discuss here
a simple version of such an equivariance under an involution
$\,\Theta:M\rightarrow M\,$ that induces a 
$\,\mathbb Z_2$-action\footnote{We shall view $\,\Zb\,$ as the
multiplicative group composed of $\,\pm1$.} on $\,M$.
Let $\,\CG=(Y,B,\CL,t)$ be a bundle gerbe over 
$\,M$. A $\,\mathbb Z_2$-equivariant extension
of $\,\CG\,$ is a gerbe $\,\tilde\CG=(\tilde Y,\tilde B,\tilde\CL,\tilde t)\,$
over $\,M\,$ such that $\,\tilde Y={\Zb}\times Y\,$
with the projection $\,\tilde Y\ni(z,y)
\mathop{\longrightarrow}\limits^{\tilde\pi}z\pi(y)\in M\,$ for $\,z=\pm1\,$ and
$\,\tilde B=p^*B\,$ for $\,p(z,y)=y$. \,We may decompose 
\qq
{\tilde Y}^{[n]}=\mathop{\sqcup}_{(z_1,\dots,z_n)}{\tilde Y}^{[n]}_{(z_1,\dots,z_n)}
\qquad{\rm for}\qquad
{\tilde Y}^{[n]}_{(z_1,\dots,z_n)}\subset
\mathop{\times}\limits_{m=1}^n\big(\{z_m\}\times Y\big).
\qqq
In particular, we may identify $\,\tilde Y^{[n]}_{1,\dots,1}\,$ with
$\,Y^{[n]}\,$ by the restriction of the map $\,p^{\times n}$. \,We demand
that under this identification,
\qq
\tilde\CL|_{{\tilde Y}^{[2]}_{(1,1)}}=\,\CL\,,
\qquad\tilde t|_{{\tilde Y}^{[3]}_{(1,1,1)}}=\,t\,.
\qqq
Finally, note that $\,\Zb\,$ acts on $\,\tilde Y\,$ by 
$\,z(z',y)=(zz',y)\,$ covering the $\,\Zb\,$ action on $\,M\,$ and this 
action lifts diagonally to $\,{\tilde Y}^{[n]}$. \,We demand that 
the $\,\Zb$-action on $\,{\tilde Y}^{[2]}\,$ lifts to a $\,\Zb$-action on 
$\,\tilde\CL\,$ by bundle isomorphisms that commute with $\,\tilde t\,$ 
and we fix such a lift. \,It is easy to see by considering the curvature 
of the line bundle $\,\tilde\CL\,$ restricted to 
$\,{\tilde Y}^{[2]}_{1,-1}\,$ that the existence of the gerbe
$\,\tilde\CG\,$ implies that the curvature form $\,H\,$ of the gerbe 
$\,\CG\,$ must be preserved by the $\,\Zb$-action on $\,M$. 
\,The notion of the $\,\Zb$-equivariant extension of a gerbe $\,\CG\,$
is equivalent to the one of the $\,\Zb$-equivariant structure on $\,\CG\,$
as defined in \cite{GSW} or \cite{GenWar}, \,see Appendix.
\vskip 0.1cm

In order to simplify the notations, we shall identify below 
$\,{\tilde Y}^{[2]}_{1,-1}\,$ with the manifold
\qq
Z=\,\big\{(y,y')\in Y\times Y\,\big|\,\pi(y)=\Theta(\pi(y'))\big\}
\label{Z}
\qqq
and the line bundle $\,\tilde\CL\,$ restricted to $\,{\tilde Y}^{[2]}_{1,-1}\,$
with a line bundle $\,\CK\,$ on $\,Z$. \,The groupoid multiplication
$\,\tilde t\,$ and the $\,\Zb\,$ symmetry of $\,\tilde\CL\,$
induce the isomorphisms
\vskip -0.2cm
\qq
\CL_{y,y'}\otimes\CK_{y',y''}\cong\CK_{y,y''}\,,\ \quad \CK_{y'',y'}
\otimes\CL_{y',y}\cong
\CK_{y'',y}\,,\ \quad\CK_{y',y''}\otimes\CK_{y'',y}\cong\CL_{y',y}\,,
\ \quad\CK_{y',y''}\cong\CK_{y'',y'}^{-1}\quad
\qqq
for $\,(y,y')\in Y^{[2]}\,$ and $\,(y',y'')\in\CZ\,$ that we shall 
abundantly use below.
\vskip 0.1cm

If $\,\Theta\,$ acts without fixed points, 
then the $\,\mathbb Z_2$-equivariant
extension $\,\tilde\CG\,$ of $\,\CG\,$ induces a gerbe
$\,\hat\CG=(\hat Y,\hat B,\hat\CL,\hat t)\,$ over the quotient
manifold $\,\hat M=M/\mathbb Z_2$,
\,see [\onlinecite{GR},\,\onlinecite{GSW}] for a discussion
of gerbes on smooth discrete quotients.
\,One takes $\,\hat Y=Y\,$ but projected to $\,\hat M\,$ rather than to
$\,M\,$ and $\,\hat B=B$. \,Then
\qq
\hat Y^{[n]}=\mathop{\sqcup}\limits_{(z_2,\dots,z_n)}
\tilde Y^{[n]}_{1,z_2,\dots,z_n}\,\subset\,\tilde Y^{[n]}
\qqq
and one sets $\,\hat\CL=\tilde\CL|_{\hat Y^{[2]}}\,$ and $\,\hat t
=\tilde t|_{\hat Y^{[3]}}$, \,the latter after the composition with the
$\,\Zb$-symmetry of $\,\tilde\CL$. \,The $\,\mathbb Z_2\,$ equivariant
extension $\,\tilde\CG\,$ will serve as the replacement for $\,\hat\CG\,$
in the case when $\,\Theta\,$ has fixed points.

\nsection{Surfaces with orientation-preserving involutions}
\label{sec:surf_invol}

\begin{figure}[b]
\begin{center}
\vskip -0.1cm
\leavevmode
\hspace{0.2cm}
        \includegraphics[width=5.7cm,height=5.6cm]{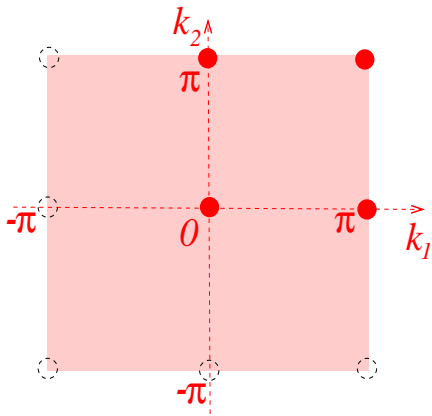}\\
        \vspace{-0.8cm} \strut
        \caption{Periodized square with the fixed
        points of the involution $\,k\mapsto -k$}
        \label{fig:TRIMs}
\end{center}
\end{figure}

\noindent Suppose that the closed oriented surface $\,\Sigma\,$
is equipped with an orientation preserving involution $\,\vartheta$.
We shall consider $\,\Sigma\,$ with the $\,\Zb$-action induced by
$\,\vartheta$. \,Let $\,\Sigma'\,$ denote the set of fixed points of
$\,\vartheta$. \,If $\,\Sigma'=\emptyset\,$ then $\,\Sigma/\mathbb Z_2\,$
is again a closed oriented surface. We shall be interested here
in the case when $\,\Sigma'\not=\emptyset$.  \,The canonical example 
will be given by the torus $\,\mathbb T^2=\mathbb R^2/(2\pi\mathbb Z^2)\,$
viewed as a square $\,[-\pi,\pi]^2\,$ with the periodic identifications of
the boundary points and the involution $\,\vartheta\,$ given by
$\,k\mapsto -k\,$ with four fixed points, \,see Fig.\,\ref{fig:TRIMs}.
The most general examples with connected $\,\Sigma\,$ and 
$\,\Sigma'\not=\emptyset\,$ are provided by the doubly ramified covers 
between Riemann surfaces 
of genus $\,g\,$ and $\,h$, \,including the hyperelliptic cover
with $\,h=0$. \,The case of $\,\mathbb T^2\,$ with the $\,k\mapsto-k\,$
involution corresponds to $\,g=1\,$ and $\,h=0$. \,The cardinality 
$\,|\Sigma'|\,$ of $\,\Sigma'\,$ satisfies the identity 
$\,|\Sigma'|=2g+2-4h\,$ following from the Riemann-Hurwitz formula. 
In particular, it is even. Around each fixed point, $\,\vartheta\,$ acts 
as $\,z\mapsto-z\,$ in an appropriate local complex coordinate.
\vskip 0.1cm

\vskip 0.1cm

If $\,\Sigma'\not=\emptyset\,$ then we shall view the quotient space
$\,\tilde\Sigma=\Sigma/\mathbb Z_2\,$ as a $\,\mathbb Z_2$-orbifold rather
than a smooth lower genus surface. As such, it possesses an orbifold
triangulation with triangles $\,\tilde c$, \,edges $\,\tilde b\,$ and vertices
$\,\tilde v$, \,the latter including the images of the fixed points 
of $\,\vartheta$ \cite{Bonahon}. The preimages of the simplices of 
that triangulation form a triangulation of $\,\Sigma\,$ with triangles 
$\,c$, \,edges $\,b\,$ and vertices $\,v$. \,The latter include the 
fixed points of $\,\vartheta\,$ whereas the other simplices of the 
triangulation of $\,\Sigma\,$ form pairs whose elements  are 
interchanged by $\,\vartheta$.

\nsection{Square root of gerbe holonomy}
\label{sec:sqrt_hol}

\noindent As we have seen, if the involution
$\,\Theta:M\rightarrow M\,$ acts without fixed points then the
$\,\Zb$-equivariant extension 
$\,\tilde\CG=(\tilde Y,\tilde B,\tilde\CL,\tilde t)\,$ of
a gerbe $\,\CG=(Y,B,\CL,t)\,$ over $\,M\,$ induces a gerbe $\,\hat\CG\,$
over $\,\hat M=M/\Zb$. \,Any map $\,\hat\phi\,$ from a closed oriented
surface $\,\hat\Sigma\,$ to the quotient manifold $\,\hat M\,$ may
be viewed as a map $\,\phi:\Sigma\rightarrow M\,$ from a double cover
$\,\Sigma\,$ of $\,\hat\Sigma\,$ to $\,M\,$ that satisfies an equivariance
condition
\qq
\phi\circ\vartheta\,=\,\Theta\circ\phi
\label{equivcond}
\qqq
for the orientation-preserving deck involution $\,\vartheta\,$ of 
$\,\Sigma\,$ interchanging the two preimages of the points of 
$\,\hat\Sigma$. \,One has the relation
\qq
\Big(\Hol_{\hat\CG}(\hat\phi)\Big)^2\,=\,\Hol_{\CG}(\phi)\,.
\qqq
The present section is devoted to a construction that 
provides an extension of such a relation to cases when
the involutions $\,\vartheta\,$ and $\,\Theta\,$ have fixed points.
We shall show that, under special conditions that will be specified below, 
a $\,\Zb$-equivariant extension $\,\tilde\CG=(\tilde Y,\tilde B,
\tilde\CL,\tilde t)\,$ of a gerbe $\,\CG=(Y,B,\CL,t)\,$ over $\,M\,$ 
permits to define a distinguished square root of the holonomy 
$\,Hol_\CG(\phi)\,$ of maps $\,\phi:\Sigma\rightarrow M\,$ satisfying
the equivariance condition (\ref{equivcond}). 
We shall construct such a square root via a local formula, a refinement
of the one for the gerbe holonomy described in Sec.\,\ref{sec:gerbe_hol}.
\vskip 0.1cm

\begin{figure}[h]
\begin{center}
\vskip 0.1cm
\leavevmode
\hspace{0.2cm}
        \includegraphics[width=4.5cm,height=4.4cm]{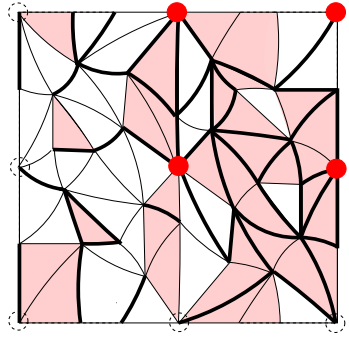}\\
        \vspace{-0.4cm} \strut
        \caption{Lift of a triangulation of $\,\mathbb T^2/\mathbb Z_2\,$
with colored selected triangles $\,c\,$ and thick selected edges $\,b$}
        \label{fig:c-and-b}
\end{center}
\end{figure}

For every triangle $\,\tilde c\,$ and every edge
$\,\tilde b\,$ of a sufficiently fine orbifold triangulation of 
$\,\tilde\Sigma=\Sigma/\Zb$,
\,we shall selects their lifts $\,c\,$ and $\,b\,$ to $\,\Sigma\,$ and then
lifts $\,s_c:c\mapsto Y\,$ and $\,s_b:b\mapsto Y\,$ 
of $\,\phi|_c\,$ and $\,\phi|_b$, \,respectively. Triangles $\,c\,$
will be considered with the orientation inherited from $\,\Sigma$. \,If
$\,\tilde b\subset\tilde c\,$ then either $\,b\subset c\,$ or $\,\vartheta(b)
\subset c$. \,An example for $\,\Sigma=\mathbb T^2\,$ and $\,\vartheta\,$
given by $\,k\mapsto-k\,$ is presented
in \,Fig.\,\ref{fig:c-and-b}.\hfill\break 

\noindent Consider now the expression 
\qq
\ee^{\ii\sum\limits_c\int_cs_c^*B}\Big(\mathop{\otimes}
\limits_{b\subset c} hol_\CL(s_c|_b,s_b)\Big)
\otimes\Big(\mathop{\otimes}\limits_{\vartheta(b)\subset c}
hol_\CK(s_c\circ\vartheta|_b,s_b)\Big),
\label{expression}
\qqq
where here and below the sums and tensor products involve only 
the selected lifts of triangles $\,\tilde c\,$ and edges $\,\tilde b$.
\,The parallel transports on the right hand side are well
defined because if $\,b\subset c\,$ then
$\,\pi\circ s_c|_b=\phi|_b=
\pi\circ s_b\,$ and if $\,\vartheta(b)\subset c\,$ then
$\,\pi\circ s_c\circ\vartheta|_b=\phi\circ\vartheta|_b=\Theta\circ\phi|_b
=\Theta\circ\pi\circ s_b$. \,Note that
\qq
&&\Big(\mathop{\otimes}\limits_{b\subset c}
hol_\CL(s|_c,s_b)\Big)\otimes\Big(\mathop{\otimes}
\limits_{\vartheta(b)\subset c}
hol_\CK(s_c\circ\vartheta|_b,s_b)\Big)\ \in\ \Big(\mathop{\otimes}
\limits_{v\in b\subset c}\CL_{s_c(v),s_b(v)}^{\pm1}\Big)
\otimes\Big(\mathop{\otimes}\limits_{v\in\vartheta(b)\subset c}\CK_{s_c(v),s_b(\vartheta(v))}^{\pm1}\Big)\cr
&&\hspace*{4.5cm}=\ \mathop{\otimes}\limits_{\tilde v}\bigg(\Big(\mathop{\otimes}
\limits_{v\in b\subset c}
\CL_{s_c(v),s_b(v)}^{\pm1}\Big)\otimes\Big(\mathop{\otimes}
\limits_{v\in\vartheta(b)\subset c}\CK_{s_c(v),s_b(\vartheta(v))}^{\pm1}\Big)\Bigg)\ \equiv\
\mathop{\otimes}\limits_{\tilde v}\CP_{\tilde v}\,,
\label{splitv}
\qqq
where on the right hand sides we regrouped together the tensor factors 
involving vertices $\,v\in\Sigma\,$ projecting to a given vertex 
$\,\tilde v\in\tilde\Sigma$. 
\vskip 0.1cm

Let us analyze the line $\,\CP_{\tilde v_0}\,$ for a fixed vertex $\,\tilde v_0$.
To this end, let us number the triangles $\,\tilde c\ni\tilde v_0\,$
counterclockwise as $\,\tilde c_1,$ $\tilde c_2$,\,\dots\,, $\tilde c_k\,$ and the
edges $\,\tilde b\ni\tilde v_0\,$ as $\,\tilde b_1$, $\tilde b_2$,\,\dots\,,
$\tilde b_k\,$ oriented towards $\,\tilde v_0$, \,starting from the edge
shared by $\,\tilde c_1\,$ and $\,\tilde c_2$, \,as \,in
\,Fig.\,\ref{fig:aroundv0} \,but now for triangles and edges of 
$\,\tilde\Sigma$. \,Let us denote by $\,v_i\,$ ($\,v'_i\,$)
\,the vertex in  $\,c_i\,$ ($\,b_i\,$) \,that projects to $\,\tilde v_0$. 
\,We shall define
\qq
\tilde s_{\tilde c_i}(\tilde v_0)\,=\,(z_i,s_{c_i}(v_i))\,\in\,\tilde Y,
\qquad\tilde s_{\tilde b_i}(\tilde v_0)\,=\,(z_i',s_{b_i}(v'_i))\,\in\,\tilde Y
\qqq
with the rules
\qq
z'_i=\begin{cases}\,z_i\qquad\ {\rm if}\ \, b_i\subset c_i\,,\quad
\ \hspace{0.13cm}i=1,\dots,k\,,\cr
(-1)z_i\ \,{\rm if}\ \,\vartheta(b_i)\subset c_i\,,\ i=1,\dots,k\,,
\end{cases}\quad
z_{i+1}=\begin{cases}\,z'_i\qquad\ \hspace{0.07cm}{\rm if}
\ \,\hspace{0.02cm}b_i\subset c_{i+1}\,,\quad\ \hspace{0.07cm}
i=1,\dots,k-1\,,\cr
\,(-1)z'_i\ \,{\rm if}\,\ \vartheta(b_i)\subset c_{i+1}\,,
\ i=1,\dots,k-1\,.
\end{cases}
\qqq
This fixes all $\,z_i\,$ and $\,z'_i\,$
except for $\,z_1\,$ whose choice will not matter.
The above choices of $\,z_i\,$ and $\,z'_i\,$ guarantee that
\qq
&&\quad\hspace{0.03cm}\tilde\CL_{\tilde s_{{\tilde c}_i}(\tilde v_0),
\tilde s_{\tilde b_i}(\tilde v_0)}\cong\begin{cases}
\CL_{s_{c_i}(v_i),s_{b_i}(v_i)}\quad\,\ {\rm if}\ \,b_i\subset c_i\,,
\ \ \ \ \ \ \,\hspace{0.1cm}i=1,\dots,k\,,\cr
\CK_{s_{c_i}(v_i),s_{b_i}(\vartheta(v_i))}\ \hspace{0.02cm}{\rm if}\ \hspace{0.03cm}
\vartheta(b_i)\subset c_i\,,\ \ \ \,i=1,\dots,k\,,\end{cases}\\ \cr
&&\tilde\CL^{-1}_{\tilde s_{{\tilde c}_{i+1}}(\tilde v_0)),
\tilde s_{\tilde b_i}(\tilde v_0)}\cong\begin{cases}
\CL^{-1}_{s_{c_i}(v_i),s_{b_i}(v_i)}\quad\ \,{\rm if}\ \, b_i\subset c_{i+1}\,,
\quad\ \hspace{0.09cm}i=1,\dots,k-1\,,\cr
\CK^{-1}_{s_{c_i}(v_i),s_{b_i}(\vartheta(v_i))}\ {\rm if}\ \,\vartheta(b_i)
\subset c_{i+1}\,,\ i=1,\dots,k-1\,.\end{cases}
\qqq
There is still the instance $\,\tilde v_0\subset\tilde b_k
\subset \tilde c_1\,$ not covered by the previous formulae. 
There are two cases here. If $\,\tilde v_0\,$
is not the image of a fixed point of $\,\vartheta\,$ then
$\,z_i=z_1\,$ if $\,v_i=v_1\,$ and $\,z_i=(-1)z_1\,$ if $\,v_i=\vartheta(v_1)$.
\,Similarly, $\,z'_i=z_1\,$ if $\,v'_i=v_1\,$ and $\,z'_i=(-1)z_1\,$
if $\,v'_i=\vartheta(v_1)$. \,Taking $\,i=k$, \,we infer that 
\qq
\tilde\CL^{-1}_{\tilde s_{\tilde c_1}(\tilde v_0),\tilde s_{\tilde b_k}(\tilde v_0)}\,
\cong\,\begin{cases}\,
\CL^{-1}_{s_{c_1}(v_1),s_{b_k}(v_1)}\qquad\,{\rm if}\quad\,b_k\subset c_1\,,\cr
\,\CK^{-1}_{s_{c_1}(v_1),s_{b_k}(\vartheta(v_1))}\quad{\rm if}\quad\vartheta(b_k)
\subset c_1\end{cases}
\qqq
in that case so that
\qq
\hspace{-0.3cm}\CP_{\tilde v_0}\cong
\tilde\CL_{\tilde s_{\tilde c_1}(\tilde v_0),\tilde s_{\tilde b_1}(\tilde v_0)}\otimes
\CL_{\tilde s_{\tilde b_1}(\tilde v_0),\tilde s_{\tilde c_2}(\tilde v_0)}\otimes
\CL_{\tilde s_{\tilde c_2}(\tilde v_0),\tilde s_{\tilde b_2}(\tilde v_0)}\otimes\,\cdots\,\otimes
\,\CL_{\tilde s_{\tilde c_k}(\tilde v_0),\tilde s_{\tilde b_k}(\tilde v_0)}\otimes
\CL_{\tilde s_{\tilde b_k}(\tilde v_0),\tilde s_{\tilde c_1}(\tilde v_0)}\cong\,\mathbb C\,,\quad
\label{27}
\qqq
where the last canonical isomorphism is obtained 
as in (\ref{6}) using the line-bundle isomorphisms $\,\tilde t\,$
of the gerbe $\,\tilde\CG$. \,The isomorphism $\,\CP_{\tilde v_0}\cong\mathbb C\,$
does not depend on the choice of the triangle $\,\tilde c_1\,$ nor on
the choice of $\,z_1\,$ due to the associativity of the groupoid
multiplication in $\,\tilde\CL\,$ and its commutation with the $\,\Zb$-action.
If, however, $\,\tilde v_0=\vartheta(v_0)\,$
for $\,v_0\in\Sigma'\,$ then $\,z'_k=z_1\,$ 
if $\,\vartheta(b_k)\subset c_1\,$
and $\,z'_k=(-1)z_1\,$ if $\,b_k\subset c_1\,$ and we have
\qq
\tilde\CL^{-1}_{(-1)\tilde s_{\tilde c_1}(\tilde v_0),\tilde s_{\tilde b_k}(\tilde v_0)}\,
\cong\,\begin{cases}\,\CL^{-1}_{s_{c_1}(v_1),s_{b_k}(v_1)}\qquad\,{\rm if}\quad\,b_k\subset c_1\,,\cr
\,\CK^{-1}_{s_{c_1}(v_1),s_{b_k}(\vartheta(v_1))}\quad{\rm if}\quad\vartheta(b_k)
\subset c_1\end{cases}
\qqq
so that
\qq
\CP_{\tilde v_0}&\cong&
\tilde\CL_{\tilde s_{\tilde c_1}(\tilde v_0),\tilde s_{\tilde b_1}(\tilde v_0)}\otimes
\CL_{\tilde s_{\tilde b_1}(\tilde v_0),\tilde s_{\tilde c_2}(\tilde v_0)}\otimes
\CL_{\tilde s_{\tilde c_2}(\tilde v_0),\tilde s_{\tilde b_2}(\tilde v_0)}\otimes\,\cdots\,\otimes
\,\CL_{\tilde s_{\tilde c_k}(\tilde v_0),\tilde s_{\tilde b_k}(\tilde v_0)}\otimes
\CL_{\tilde s_{\tilde b_k}(\tilde v_0),(-1)\tilde s_{\tilde c_1}(\tilde v_0)}\cr
&\cong&\tilde\CL_{\tilde s_{\tilde c_1}(\tilde v_0),(-1)\tilde s_{\tilde c_1}(\tilde v_0)}\,\cong\,
\tilde\CL_{(1,s_{c_1}(v_0)),(-1,s_{c_1}(v_0))}\,=\,\CK_{s_{c_1}(v_0),s_{c_1}(v_0)}
\label{Pzfp}
\qqq
in this case. Can we canonically trivialize the latter lines? \,Let
\qq
M'\,=\,\big\{x\in M\,\big|\,\Theta(x)=x\big\} 
\qqq
be the set of fixed points of $\,\Theta\,$ that, \,for simplicity, 
\,we shall assume to be a submanifold of $\,M$. 
\,Let $\,Y'=\pi^{-1}(M')\subset Y$. \,Note that
the equivariance (\ref{equivcond}) implies that $\,\phi({\Sigma'})
\subset M'\,$
so that $\,s_{c_1}(v_0)\in Y'$. \,Consider the map
\qq
Y'\ni y'\,\mathop{\longrightarrow}\limits^r\,((1,y'),(-1,y'))\,
\in \tilde Y^{[2]}
\qqq
and the flat line bundle $\,\CN'=r^*\tilde\CL$.
\,Note that (\ref{Pzfp}) may be rewritten as the relation
\qq
\CP_{\tilde v_0}\,\cong\,\CN'_{s_{c_1}(v_0)}\,.
\label{CPsc1}
\qqq
What is easy to see is that the square of the line bundle $\,\CN'\,$ 
possesses a natural trivialization
\qq
\CN'^{\,2}\,\cong\,Y'^{[2]}\times\mathbb C
\label{trivCN'2}
\qqq
given on the fibers by the $\,\Zb$-action on $\,\tilde\CL\,$ and its
groupoid multiplication $\,\tilde t\,$:
\qq
\CN'^{\,2}_{y'}=\tilde\CL^{\,2}_{(1,y'),(-1,y')}\cong\tilde\CL_{(1,y'),(-1,y')}\otimes
\tilde\CL_{(-1,y'),(1,y')}\cong\tilde\CL_{(1,y'),(1,y')}\cong\mathbb C\,.
\qqq
Denote by $\,\pi'\,$ the restriction
of the surbmesion $\,\pi\,$ and by $\,B'\,$ the restriction of the 2-form
$\,B\,$ to $\,Y'$. \,The map $\,\pi':Y'\rightarrow M'\,$ is a surjective
submersion and $\,\tilde Y'^{[2]}\,$ may be identified with a submanifold
of $\,Y^{[2]}\,$. It makes then sense to consider the line bundles
$\,\CL'=\CL|_{Y'^{[2]}}\,$ with the groupoid multiplication $\,t'\,$
induced from the one of $\,\CL$. \,There is a natural isomorphism
of line bundles over $\,Y'^{[2]}$
\qq
\CL'\otimes p'^*_2\CN'\,\mathop{\longrightarrow}\limits^{\nu'}\,
p'^*_1\CN'\otimes\CL'
\label{nu'}
\qqq
given again by the groupoid multiplication $\,\tilde t\,$ and 
the $\,\mathbb Z_2$-action on $\,\tilde\CL$. \,Indeed, fiber-wise, 
\qq
&&\CL'_{y'_1,y'_2}\otimes\CN'_{y'_2}\,\cong\,\tilde\CL_{(1,y'_1),(1,y'_2)}\otimes
\tilde\CL_{(1,y'_2),(-1,y'_2)}\cr
&&\cong\,\tilde\CL_{(1,y'_1),(-1,y'_2)}\,\cong\,\tilde\CL_{(1,y'_1),(-1,y'_1)}
\otimes\tilde\CL_{(-1,y'_1),(-1,y'_2)}\,\cong\CN'_{y'_1}\otimes\CL_{y'_1,y'_2}
\label{nu'1}
\qqq
which commutes with the groupoid multiplication in $\,\CL'$, \,i.e. such
that for $\,(y'_1,y'_2,y'_3)\in Y'^{[3]}\,$ the isomorphism of lines
\qq
\CL'_{y'_1,y'_2}\otimes\CL'_{y'_2,y'_3}\otimes\CN'_{y'_3}
\mathop{\longrightarrow}\limits^{t'\otimes\Id}\CL'_{y'_1,y'_3}\otimes\CN'_{y'_3}
\mathop{\longrightarrow}\limits^{\nu'}\,\CN'_{y'_1}\otimes\CL_{y'_1,y'_3}
\qqq
coincides with
\qq
\CL'_{y'_1,y'_2}\otimes\CL'_{y'_2,y'_3}\otimes\CN'_{y'_3}
\mathop{\longrightarrow}\limits^{\Id\otimes\nu'}\,
\CL'_{y'_1,y'_2}\otimes\CN'_{y'_2}\otimes\CL'_{y'_2,y'_3}
\mathop{\longrightarrow}\limits^{\nu'\otimes\Id}\,
\CN'_{y'_1}\otimes\CL_{y'_1,y'_2}\otimes\CL'_{y'_2,y'_3}
\mathop{\longrightarrow}\limits^{\Id\otimes t'}
\,\CN'_{y'_1}\otimes\CL'_{y'_1,y'_3}\,.\quad
\label{19}
\qqq
The isomorphism $\,\nu'\,$ allows to canonically identify the lines
$\,\CN'_{y'}\,$ for all $\,y'\,$ over the same point $\,x\in M'\,$
and the bundle $\,\CN'\,$ with a pullback $\,\pi^*N'\,$ of a flat bundle
$\,N'\,$ over $\,M'$. \,A straightforward check shows that
the trivialization (\ref{trivCN'2}) commutes
with $\,\nu'^2\,$ so that it defines a trivialization of the flat
line bundle $\,N'^2\,$ over $\,M'$. \,In general, that does not
imply the trivializability of the flat line bundle $\,N'$. \,If, however,
$\,M'\,$ is simply connected then $\,N'\,$ is trivializable (as any
flat line bundle over a simply connected manifold) and we may choose
its trivialization so that it squares to the trivialization of $\,N'^2\,$
induced by (\ref{trivCN'2}). Besides, if $\,M'\,$ is also connected
then such a trivialization of $\,N'\,$ is defined up to a global sign.
It induces a preferred trivialization of $\,\CN'\,$ also defined
modulo a global sign. Such a trivialization allows to identify
the lines $\,\CP_{\tilde v_0}\,$ of (\ref{Pzfp}) with $\,\mathbb C$,
\,again up to a global sign. Let us check that the above identification 
does not depend on the choice of the initial triangle 
$\,\tilde c_1\ni\tilde v_0$. \,The choice of $\,\tilde c_2\,$ as the 
initial triangle gives
\qq
\CP_{\tilde v_0}&\cong&
\CL_{\tilde s_{\tilde c_2}(\tilde v_0),\tilde s_{\tilde b_2}(\tilde v_0)}\otimes\,\cdots\,\otimes
\,\CL_{\tilde s_{\tilde c_k}(\tilde v_0),\tilde s_{\tilde b_k}(\tilde v_0)}\cr
&&\otimes\CL_{\tilde s_{\tilde b_k}(\tilde v_0),(-1)\tilde s_{\tilde c_1}(\tilde v_0)}
\otimes\tilde\CL_{(-1)\tilde s_{\tilde c_1}(\tilde v_0),(-1)\tilde s_{\tilde b_1}(\tilde v_0)}\otimes
\CL_{(-1)\tilde s_{\tilde b_1}(\tilde v_0),(-1)\tilde s_{\tilde c_2}(\tilde v_0)}
\cr
&\cong&\tilde\CL_{\tilde s_{\tilde c_2}(\tilde v_0),(-1)\tilde s_{\tilde c_2}(\tilde v_0)}\,\cong\,
\tilde\CL_{(1,s_{c_2}(v_0)),(-1,s_{c_2}(v_0))}
\label{Pzfp1}
\qqq
i.e.
\qq
\CP_{\tilde v_0}\,\cong\,\CN'_{s_{c_2}(v_0)}
\label{CPsc2}
\qqq
The isomorphisms (\ref{CPsc1}) and (\ref{CPsc2}) may be summarized as
resulting from the ones
\qq
&\tilde\CL_{\tilde s_{\tilde c_1}(\tilde v_0),\tilde s_{\tilde c_2}(\tilde v_0)}
\otimes\CL_{\tilde s_{\tilde c_2}(\tilde v_0),(-1)\tilde s_{\tilde c_1}(\tilde v_0)}
\,\cong\,\tilde\CL_{\tilde s_{\tilde c_1}(\tilde v_0),(-1)\tilde s_{\tilde c_1}(\tilde v_0)}\,,&
\label{1stisom}\\
&\tilde\CL_{\tilde s_{\tilde c_2}(\tilde v_0),(-1)\tilde s_{\tilde c_1}(\tilde v_0)}\otimes
\tilde\CL_{(-1)\tilde s_{\tilde c_1}(\tilde v_0),(-1)\tilde s_{\tilde c_2}(\tilde v_0)}\,\cong\,
\tilde\CL_{\tilde s_{\tilde c_2}(\tilde v_0),(-1)\tilde s_{\tilde c_2}(\tilde v_0)}\,,&
\label{2ndisom}
\qqq
respectively. Tensoring the both sides of (\ref{1stisom})
with $\,\tilde\CL_{(-1)\tilde s_{\tilde c_1}(\tilde v_0),
(-1)\tilde s_{\tilde c_2}(\tilde v_0)}\,$ and the both sides of (\ref{2ndisom})
with $\,\tilde\CL_{\tilde s_{\tilde c_1}(\tilde v_0),\tilde s_{\tilde c_2}(\tilde v_0)}$,
\,we make the left hands equal whereas the identification of the 
right hand sides agrees with the interpretation of the 
line bundle $\,\CN'\,$ as the pullback of the line bundle $\,N'$,
\,see (\ref{nu'}) and (\ref{nu'1}). We infer that (\ref{CPsc1}) and
(\ref{CPsc2}) induce the same isomorphisms of lines
\qq
\CP_{\tilde v_0}\,\cong\,N'_{\phi(v_0)}
\label{CPN'}
\qqq
which is then independent of the choice of the initial triangle.
If $\,M'\,$ a 1-connected then, \,using the trivialization
of $\,N'\,$ described above and defined modulo a global sign, we
obtain an isomorphism $\,\CP_{\tilde v_0}\cong\mathbb C\,$ defined
up to a sign that is the same for all fixed points $\,v_0\,$ of $\,\vartheta$.
Since the number of such fixed points is even, this sign ambiguity disappears
when we take the tensor product of such identifications over all $\tilde v_0$.
\vskip 0.1cm

Summarizing the above discussion, we conclude that if $\,M'\,$ is
a 1-connected submanifold of $\,M\,$ then the expression (\ref{expression})
may be identified with a phase\footnote{The parallel transports
occurring in (\ref{expression}) and the isomorphisms
$\,\CP_{\tilde v}\cong\mathbb C\,$ preserve the hermitian structures.}
in $\,U(1)\,$ that is independent of the sign in the choice of
the trivialization of $\,N'$.
\vskip 0.4cm

\noindent{\bf Proposition 3.} \ The $\,U(1)$-phase associated to the expression 
(\ref{expression}) is independent of the choice of maps $\,s_c\,$
and $\,s_b\,$ lifting $\,\phi|_c\,$ and $\,\phi|_b\,$ to $\,Y$, \,of the lifts 
$\,c\,$ and $\,b\,$ of simplices $\,\tilde c\,$ and $\,\tilde b\,$ to
$\,\Sigma\,$  and of the orbifold triangulation of $\,\tilde\Sigma$. 
\vskip 0.3cm

\noindent{\bf Proof.} \,We shall proceed similarly as the the proof
of Proposition 1.
\vskip 0.1cm

1.\,\ If we change the map $\,s_{c_0}\,$ to $\,s'_{c_0}\,$ for the lift
$\,c_0\,$ of a triangle $\,\tilde c_0\,$ then
\qq
&&\ee^{\ii\int_{c_0}s'^*_{c_0}B}\,=\,\ee^{\ii\int_{c_0} s_{c_0}^*B}\,
\ee^{\ii\int_{c_0}(s'^*_{c_0}B-s^*_{c_0}B)}\,=\,\ee^{\ii\int_{c_0}s_{c_0}^*B}\,
\ee^{\ii\int_{c_0}(s_{c_0},s'_{c_0})^*F_{\CL}}\cr\cr
&&\cong\,\ee^{\ii\int_{c_0}s_{c_0}^*B}
\Big(\mathop{\otimes}\limits_{b\subset c_0}hol_{\CL}
(s_{c_0}|_b,s'_{c_0}|_b)\Big)\otimes\Big(\mathop{\otimes}
\limits_{\vartheta(b)\subset c_0}hol_{\CL}
(s_{c_0}|_{\vartheta(b)},s'_{c_0}|_{\vartheta(b)})\Big)\cr
&&=\,\ee^{\ii\int_{c_0} s_{c_0}^*B}
\Big(\mathop{\otimes}\limits_{b\subset c_0}hol_{\CL}
(s_{c_0}|_b,s'_{c_0}|_b)\Big)\otimes\Big(\mathop{\otimes}
\limits_{\vartheta(b)\subset c_0}hol_{\CL}
(s_{c_0}\circ\vartheta|_b,s'_{c_0}\circ\vartheta|_b)\Big).\quad
\label{11}
\qqq
Using the relations
\qq
&\mathop{\otimes}\limits_{b\subset c_0}\Big(hol_{\CL}
(s_{c_0}|_b,s'_{c_0}|_b)\otimes hol_{\CL}(s'_{c_0}|_b,s_b)\Big)
\,\mathop{\cong}\limits^t\,
\mathop{\otimes}\limits_{b\subset c_0}hol_{\CL}(s_{c_0}|_b,s_b)\,,&\cr
&\mathop{\otimes}\limits_{\vartheta(b)\subset c_0}
\Big(hol_{\CL}(s_{c_0}\circ\vartheta|_b,s'_{c_0}\circ\vartheta|_b)
\otimes hol_{\CK}(s'_{c_0}\circ\vartheta|_b,s_b)\Big)\,\mathop{\cong}
\limits^{\tilde t}\,\mathop{\otimes}\limits_{\vartheta(b)\subset c_0}
hol_{\CK}(s_{c_0}\circ\vartheta|_b,s_b)\,,&
\qqq
one shows that the expression (\ref{expression})
after the change is equivalent to the one before the change 
and the associativity of $\,\tilde t\,$ guarantees that both 
define the same phase if $\,M'\,$ is a 1-connected submanifold of $\,M$.
\vskip 0.1cm

If we change the lift $\,c_0\,$ of a triangle $\,\tilde c_0\,$ to
$\,c_0'=\vartheta(c_0)\,$ and the map $\,s_{c_0}\,$ to $\,s_{c'_0}\,$
then
\qq
&&\ee^{\ii\int_{c'_0}s_{c'_0}^*B}=\ee^{\ii\int_{c_0}s_{c_0}^*B}\,
\ee^{\ii\int_{c_0}((s_{c'_0}\circ\vartheta)^*B-s_{c_0}^*B)}=\ee^{\ii\int_{c_0}s_{c_0}^*B}\,
\ee^{\ii\int_{c_0}(s_{c_0},s_{c'_0}\circ\vartheta)^*F_\CK}\cr\cr
&&\cong\ee^{\ii\int_{c_0}s_{c_0}^*B}\,\Big(\mathop{\otimes}\limits_{b\subset c_0}
hol_\CK(s_{c_0}|_b,s_{c'_0}\circ\vartheta|_b)
\Big)\otimes\Big(\mathop{\otimes}\limits_{\vartheta(b)\subset c_0}
hol_\CK(s_{c_0}|_{\vartheta(b)},s_{c'_0}\circ\vartheta|_{\vartheta(b)})\Big)\cr
&&=\ee^{\ii\int_{c_0}s_{c_0}^*B}\,\Big(\mathop{\otimes}\limits_{b\subset c_0}
hol_\CK(s_{c_0}|_b,s_{c'_0}\circ\vartheta|_b)
\Big)\otimes\Big(\mathop{\otimes}\limits_{\vartheta(b)\subset c_0}
hol_\CK(s_{c_0}\circ\vartheta|_b,
s_{c'_0}|_b)\Big).
\qqq
Using the relations
\qq
&&\Big(\mathop{\otimes}\limits_{b\subset c_0}
hol_\CK(s_{c_0}|_b,s_{c'_0}\circ\vartheta|_b)
\Big)\otimes\Big(\mathop{\otimes}\limits_{\vartheta(b)\subset c'_0}
hol_\CK(s_{c'_0}\circ\vartheta|_b,s_b)\Big)\cr
&&=\,\mathop{\otimes}\limits_{b\subset c_0}
\Big(hol_\CK(s_{c_0}|_b,s_{c'_0}\circ\vartheta|_b)
\otimes hol_\CK(s_{c'_0}\circ\vartheta|_b,s_b)\Big)\,
\mathop{\cong}\limits^{\tilde t}\,\mathop{\otimes}\limits_{b\subset c_0}
hol_\CL(s_{c_0}|_b,s_b)\,,\cr\cr
&&\Big(\mathop{\otimes}\limits_{\vartheta(b)\subset c_0}
hol_\CK(s_{c_0}\circ\vartheta|_b,
s_{c'_0}|_b)\Big)\otimes\Big(\mathop{\otimes}\limits_{b\subset c'_0}
hol_\CL(s_{c'_0}|_b,s_b)\Big)\cr
&&=\,\mathop{\otimes}\limits_{\vartheta(b)\subset c_0}
\Big(hol_\CK(s_{c_0}\circ\vartheta|_b,s_{c'_0}|_b)\otimes hol_\CL(s_{c'_0}|_b,s_b)
\Big)\,\mathop{\cong}\limits^{\tilde t}\,\mathop{\otimes}\limits_{\vartheta(b)
\subset c_0}hol_\CK(s_{c_0}\circ\vartheta|_b,s_b)
\qqq
one shows that, again, the expression (\ref{expression})
after the change is equivalent to the one before the change and, as before, 
the associativity of $\,\tilde t\,$ guarantees that both define the 
same phase.
\vskip 0.1cm

2.\ \,Similarly, if we change the map $\,s_{b_0}\,$ to $\,s'_{b_0}\,$
for the lift $\,b_0\,$ of an  edge $\,\tilde b_0\,$ then
\qq
&&\qquad\ \,\,\mathop{\otimes}\limits_{c\supset b_0}
hol_\CL(s_c|_{b_0},s'_{b_0})\,\mathop{\cong}\limits^t\,
\mathop{\otimes}\limits_{c\supset b_0}
\Big(hol_\CL(s_c|_{b_0},s_{b_0})\otimes hol_\CL(s_{b_0},s'_{b_0}))\Big),\\
&&\mathop{\otimes}\limits_{c\supset\vartheta(b_0)}
hol_\CK(s_c\circ\vartheta|_{b_0},s'_{b_0})\,\mathop{\cong}\limits^{\tilde t}\,
\mathop{\otimes}\limits_{c\supset\vartheta(b_0)}
\Big(hol_\CK(s_c\circ\vartheta|_{b_0},s_{b_0})\otimes
hol_\CL(s_{b_0},s'_{b_0}))\Big).
\qqq
But
\vskip -0.5cm
\qq
\Big(\mathop{\otimes}\limits_{c\supset b_0}hol_\CL(s_{b_0},s'_{b_0})\Big)
\otimes\Big(\mathop{\otimes}\limits_{c\supset\vartheta(b_0)}
hol_\CL(s_{b_0},s'_{b_0})\Big)\,\cong\,1
\qqq
as the terms appear in dual pairs corresponding to two triangles $\,\tilde c\,$
bordering the same edge $\,\tilde b_0\,$ that induce on it and on the
corresponding edge $\,b_0\,$ opposite orientations.
\vskip 0.1cm

If we change the lift $\,b_0\,$ of an edge $\,\tilde b_0\,$ to
$\,b_0'=\vartheta(b_0)\,$ and the map $\,s_{b_0}\,$ to $\,s_{b'_0}\,$
then
\vskip -0.5cm
\qq
&&\hspace{-0.4cm}\mathop{\otimes}\limits_{c\supset b'_0}hol_\CL(s_c|_{b'_0},s_{b'_0})
\,=\hspace{-0.1cm}\mathop{\otimes}\limits_{c\supset\vartheta(b_0)}\hspace{-0.1cm}
hol_\CL(s_c\circ\vartheta|_{b_0},s_{b'_0}\circ\vartheta)\,
\mathop{\cong}\limits^{\tilde t}\hspace{-0.1cm}
\mathop{\otimes}\limits_{c\supset\vartheta(b_0)}\hspace{-0.1cm}
\Big(hol_\CK(s_c\circ\vartheta|_{b_0},s_{b_0})\otimes hol_\CK(s_{b_0},s_{b'_0}
\circ\vartheta)\Big),\quad\\
&&\hspace{-0.6cm}\mathop{\otimes}\limits_{c\supset\vartheta(b'_0)}
\hspace{-0.1cm}hol_\CK(s_c\circ\vartheta|_{b'_0},s_{b'_0})\,=\,
\mathop{\otimes}\limits_{c\supset b_0}
hol_\CK(s_c|_{b_0},s_{b'_0}\circ\vartheta)
\,\mathop{\cong}\limits^{\tilde t}
\mathop{\otimes}\limits_{c\supset b_0}\hspace{-0.1cm}
\Big(hol_\CL(s_c|_{b_0},s_{b_0})\otimes
hol_\CK(s_{b_0},s_{b'_0}\circ\vartheta))\Big).
\qqq
However,
\vskip -0.6cm
\qq
\Big(\mathop{\otimes}\limits_{c\supset\vartheta(b_0)}
hol_\CK(s_{b_0},s_{b'_0}\circ\vartheta)\Big)\otimes\Big(\mathop{\otimes}
\limits_{c\supset b_0}hol_\CK(s_{b_0},s'_{b_0}\circ\vartheta)\Big)\,\cong\,1
\qqq
as, again, the terms appear in dual pairs corresponding to two triangles
$\,\tilde c\,$ bordering the same edge $\,\tilde b_0\,$ and inducing on
$\,b_0\,$ opposite orientations.
\vskip 0.1cm

3.\ \,The independence of the phase associated to (\ref{expression})
on the orbifold triangulation of $\,\tilde\Sigma\,$ is proven by
using the two Pachner moves depicted on \,Fig.\,\ref{fig:subdiv}
\,for that triangulation and the corresponding subdivisions of
the lifted triangles and edges. The argument that such moves
lead to equivalent expressions (\ref{expression}) is then
essentially the same as in the proof of Proposition 1.
\vskip -0.1cm

\hspace{14cm}$\blacksquare$
\vskip 0.3cm

We are now ready to define the square root of the gerbe holonomy of 
equivariant maps.
\vskip 0.4cm

\noindent{\bf Definition 2.} \ Suppose that $\,M\,$ is a manifold
with a $\,\Zb$-action induced by an involution $\,\Theta\,$ 
with the fixed-point set $\,M'\,$ that is a 1-connected
submanifold of $\,M$. \,Let $\,\tilde\CG\,$ be a $\,\Zb$-equivariant extension 
of a gerbe $\,\CG\,$ over $\,M\,$ and $\,\phi:\Sigma\rightarrow M\,$
be a map satisfying the equivariance condition (\ref{equivcond})
for an orientation-preserving involution $\,\vartheta:\Sigma\rightarrow\Sigma\,$
with discrete fixed points. Then we set
\qq
\sqrt{\Hol_\CG(\phi)}\,=\,\ee^{\ii\sum\limits_c\int_cs_c^*B}\Big(\mathop{\otimes}
\limits_{b\subset c}hol_\CL(s_c|_b,s_b)\Big)
\otimes\Big(\mathop{\otimes}\limits_{\vartheta(b)\subset c}
hol_\CK(s_c\circ\vartheta|_b,s_b)\Big),
\label{sqrtHol}
\qqq
where the right hand side is identified with a $\,U(1)$-phase the way
described above.
\vskip 0.3cm

It remains to show
\vskip 0.3cm

\noindent{\bf Lemma 1.}
\vskip -0.5cm
\qq
\Big(\sqrt{\Hol_\CG(\phi)}\Big)^2\,=\,Hol_\CG(\phi)\,.
\label{lemma1}
\qqq
\vskip 0.3cm

\noindent{\bf Proof of Lemma 1.} \ Recall that on the right hand side
of (\ref{sqrtHol}), $\,c\,$ and $\,b\,$ run over the selected lifts 
of simplices $\,\tilde c\,$ and $\,\tilde b\,$ of the orbifold 
triangulation of $\,\tilde\Sigma$. \,By Proposition 2, \,we may also 
use on the right hand side of (\ref{sqrtHol})
the opposite choices $\,c'=\vartheta(c),\,b'=\vartheta(b)\,$ of such lifts.
\,Now
\qq
&&\ee^{\ii\sum\limits_c\int_cs_c^*B}\Big(\mathop{\otimes}
\limits_{b\subset c}hol_\CL(s_c|_b,s_b)\Big)
\otimes\Big(\mathop{\otimes}\limits_{\vartheta(b)\subset c}
hol_\CK(s_c\circ\vartheta|_b,s_b)\Big)\cr
&&=\,\ee^{\ii\sum\limits_c\int_cs_c^*B}\Big(\mathop{\otimes}
\limits_{b\subset c}hol_\CL(s_c|_b,s_b)\Big)
\otimes\Big(\mathop{\otimes}\limits_{b'\subset c}
hol_\CK(s_c|_{b'},s_b\circ\vartheta)\Big)\cr
&&\mathop{\cong}\limits^{\tilde t}\,\,\ee^{\ii\sum\limits_c\int_cs_c^*B}
\Big(\mathop{\otimes}\limits_{b\subset c}
hol_\CL(s_c|_b,s_b)\Big)\otimes\Big(\mathop{\otimes}
\limits_{b'\subset c}\big(hol_\CL(s_c|_{b'},s_{b'})
\otimes hol_\CK(s_{b'},s_b\circ\vartheta)\big)\Big).
\qqq
Similarly
\qq
&&\ee^{\ii\sum\limits_{c'}\int_{c'}s_{c'}^*B}\Big(\mathop{\otimes}
\limits_{b'\subset c'}hol_\CL(s_{c'}|_{b'},s_{b'})\Big)
\otimes\Big(\mathop{\otimes}\limits_{\vartheta(b')\subset c'}
hol_\CK(s_{c'}\circ\vartheta|_{b'},s_{b'})\Big)\cr
&&=\,\ee^{\ii\sum\limits_{c'}\int_{c'}s_{c'}^*B}\Big(\mathop{\otimes}
\limits_{b'\subset c'}hol_\CL(s_{c'}|_{b'},s_{b'})\Big)
\otimes\Big(\mathop{\otimes}\limits_{b\subset c'}
hol_\CK(s_{c'}|_b,s_{b'}\circ\vartheta)\Big)\cr
&&\mathop{\cong}\limits^{\tilde t}\,\,\ee^{\ii\sum\limits_{\tilde c}\int_{c'}s_{c'}^*B}
\Big(\mathop{\otimes}\limits_{b'\subset c'}
hol_\CL(s_{c'}|_{b'},s_{b'})\Big)\otimes\Big(\mathop{\otimes}
\limits_{b\subset c'}\big(hol_\CL(s_{c'}|_b,s_b)\otimes
hol_\CK(s_b,s_{b'}\circ\vartheta)\big)\Big)\cr
&&=\,\ee^{\ii\sum\limits_{c'}\int_{c'}s_{c'}^*B}
\Big(\mathop{\otimes}\limits_{b'\subset c'}
hol_\CL(s_{c'}|_{b'},s_{b'})\Big)\otimes\Big(\mathop{\otimes}
\limits_{b\subset c'}\big(hol_\CL(s_{c'}|_b,s_b)\otimes
hol_\CK(s_b\circ\vartheta,s_{b'})\big)\Big).
\qqq
Using the relation
\qq
\mathop{\otimes}\limits_{b'\subset c}
\Big(hol_\CK(s_{b'},s_b\circ\vartheta)\otimes
hol_\CK(s_b\circ\vartheta,s_{b'})\Big)
\,\mathop{\cong}\limits^{\tilde t}\,1
\qqq
(involving also the $\,\Zb$-symmetry of line bundle $\,\tilde\CL$),
\,we infer that the tensor product of the two versions of the right
hand side of (\ref{sqrtHol}) is equivalent to
\qq
&&\quad\ee^{\ii\sum\limits_{\tilde c}(\int_cs_c^*B+\int_{c'}s_{c'}^*B)}
\Big(\mathop{\otimes}\limits_{b\subset c}
hol_\CL(s_c|_b,s_b)\Big)\otimes\Big(\mathop{\otimes}
\limits_{b'\subset c}hol_\CL(s_c|_{b'},s_{b'})\Big)\cr
&&\hspace{2.85cm}
\otimes\,\Big(\mathop{\otimes}
\limits_{b'\subset c'}hol_\CL(s_{c'}|_{b'},s_{b'})\Big)
\otimes\Big(\mathop{\otimes}\limits_{b\subset c'}hol_\CL(s_{c'}|_b,s_b)\Big)
\qqq
which is a version of the right hand side of (\ref{HolCG}) for the
triangulation of $\,\Sigma\,$ induced from that of $\,\tilde\Sigma$.
A straightforward (although somewhat tedious) check shows that the above
identifications commute with the ones associating $\,U(1)$-phases to the right
hand sides of the two versions of (\ref{sqrtHol}) and to (\ref{HolCG}).
This proves the identity (\ref{lemma1}).
\vskip -0.15cm

\hspace{13cm}$\square$
\vskip 0.4cm

\noindent{\bf Example 2.} \ If $\,\phi:\Sigma\mapsto M\,$ is constant with the
value $\,m'\in\,M'\,$ then $\,\sqrt{\Hol_\CG(\phi)}=1$. \,This is easily
shown choosing all $\,s_c$ and $\,s_b\,$ involved in
the expression on the right hand side of (\ref{sqrtHol}) constant and taking
the same value $\,y'\in Y'$. \,Then the right hand side of (\ref{sqrtHol})
is equal to $\,1\,$ as an element of the line
\qq
\Big(\mathop{\otimes}
\limits_{v\in b\subset c}\CL_{y',y'}^{\pm1}\Big)
\otimes\Big(\mathop{\otimes}\limits_{v\in\vartheta(b)\subset c}
\CK_{y',y'}^{\pm1}\Big)\,\cong\,\mathbb C\,,
\label{linects}
\qqq
see (\ref{splitv}), if the last isomorphism results from the fact that each
line is accompanied by its dual corresponding to the opposite end of $\,b$.
\,But also for the individual lines one has
the canonical isomorphisms $\,\CL_{y',y'}^{\pm1}\cong\mathbb C\,$ 
and $\,\CK_{y',y'}^{\pm1}=\CN'^{\pm 1}_{y'}\cong\mathbb C\,$ (in the last case
up to a sign) and the latter isomorphisms agree with the ones resulting in
the identifications $\,\CP_{\tilde v}\cong\mathbb C\,$ on which the interpretation
of the right hand side of (\ref{sqrtHol}) as an $\,U(1)$-phase is based.
Clearly the two identifications of the line (\ref{linects}) with
$\,\mathbb C\,$ also agree proving the announced equality.
\vskip 0.4cm

If $\,\CD:\Sigma\rightarrow\Sigma\,$ is a diffeomorphism that commutes
with $\,\vartheta\,$ and preserves or reverses the orientation then,
respectively,
\qq
\sqrt{\Hol_\CG(\phi)}\,=\,\sqrt{\Hol_\CG(\phi\circ\CD)}^{\hspace{0.02cm}\pm1},
\label{sqrtD}
\qqq
as may be easily seen by computing the left and the right hand sides using,
the triangulations, the lifts $\,c,\,b\,$ and the maps $\,s_c,\,s_b\,$ related
by $\,\CD$.

\nsection{Homotopic formula for the square root of gerbe holonomy}
\label{sec:homot_form}

\begin{figure}[h]
\vskip 0.1cm
\begin{center}
\leavevmode
 \includegraphics[width=4.6cm,height=3.2cm]{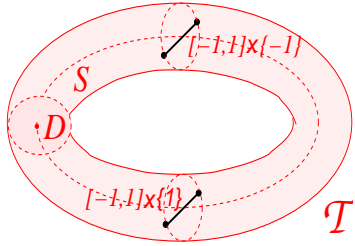}
\vskip -0.1cm
\caption{$\CT=D\times S\,$ with the fixed-point set $\,\CT'\,$ 
of the involution $\,\zeta:(z,v)\mapsto(\bar z,\bar v)$}
\label{fig:tildeSigma}
\end{center}
\end{figure}

\noindent Let $\,\CT\,$ be a compact oriented
3-manifold with boundary $\,\partial\CT=\Sigma\,$ equipped
with an orientation-preserving involution $\,\zeta\,$
reducing to $\,\vartheta\,$ on the boundary. We shall assume that at the fixed
points of $\,\zeta\,$ its derivative has one eigenvalue $\,1\,$ and two
eigenvalues $\,-1$. \,Then fixed-point set of $\,\zeta\,$ forms necessarily 
a $\,1d\,$ submanifold with boundary $\,\CT'\subset\CT\,$ such that 
$\,\partial\CT'=\Sigma'$. \,An example for $\,\Sigma=\mathbb T^2=\mathbb R^2/
(2\pi\mathbb Z^2)\,$ with the $\,k\mapsto-k\,$ involution would be 
$\,\CT=D\times S$, \,where $\,D\,$ is the unit disc and $\,S\,$
the unit circle in the complex plane, with the boundary identification
induced by the map
$\,(k_1,k_2)\mapsto(\ee^{\ii k_1},\ee^{\ii k_2})\,$ and with $\,\zeta\,$
given by $\,(z,v)\mapsto(\bar z,\bar v)$. \,In this case, $\,\CT'
=[-1,1]\times\{1\}\cup[-1,1]\times\{-1\}$, \,see Fig.\,\ref{fig:tildeSigma}.
\vskip 0.1cm

\noindent{\bf Proposition 4.} \ Suppose that a map
$\,\phi:\Sigma\rightarrow M\,$ satisfying (\ref{equivcond}) has
an extension $\,\psi:\CT\rightarrow M\,$ such that
\qq
\psi\circ\zeta\,=\,\Theta\circ\psi\,.
\label{equivcond2}
\qqq
Assume that the fixed-point set $\,M'\m$ of $\,\Theta\,$ is a 1-connected
submanifold of $\,M$. \,Then for any $\,\Zb$-equivariant extension 
$\,\tilde\CG\,$ of a gerbe $\,\CG\,$ on $\,M\,$ with curvature $\,H$,
\qq
\sqrt{\Hol_\CG(\phi)}\ =\ \ee^{\frac{\ii}{2}\int\limits_{\CT}\psi^*H}\,.
\label{sqrtSWZ}
\qqq
In particular, given $\,\phi$, \,the right hand side does not depend on
its equivariant extension $\,\psi$.
\vskip 0.3cm

\begin{figure}[h]
\vskip -0.1cm
\begin{center}
\leavevmode
 \includegraphics[width=4.6cm,height=3.5cm]{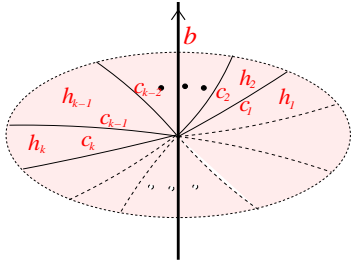}
\vskip -0.1cm
\caption{Simplices $\,h,\,c\,$ containing an edge 
$\,b\subset\CT'\,$ - a perpendicular cut. $\,\zeta\,$ is the rotation 
by $\,\pi\,$ around $\,b$}
\label{fig:aroundedge}
\end{center}
\vskip -0.03cm
\end{figure}

\noindent{\bf Proof.} \ The involution $\,\zeta\,$ generates on $\,\CT\,$
a $\,\Zb$-action and we shall view the quotient space 
$\,\tilde\CT=\CT/\Zb\,$
as a $\,\Zb$-orbifold. Let us fix a sufficiently 
fine orbifold triangulation
of $\,\tilde\CT\,$ \cite{Bonahon} with simplices
$\,\tilde h,\tilde c,\tilde b,\tilde v$. \,By definition, its restrictions 
to the 2-dimensional boundary $\,\tilde\Sigma=\Sigma/\Zb\,$ 
and to the 1-dimensional
fixed-point set $\,\CT'\,$ induce triangulations of the latter. The
preimages of simplices of the orbifold triangulation give rise to a
triangulation of $\,\CT\,$ with simplices 
permuted within pairs by $\,\zeta$,
\,except for the ones in $\,\CT'\,$ that are left invariant.
Let us fix for the simplices $\,\tilde h,\,\tilde c,\,\tilde b\,$ their lifts
$\,h,\,c,\,b\,$ to $\,\CT\,$ and the maps $\,s_h:h\rightarrow Y$,
$\,s_c:c\rightarrow Y$, $\,s_b:b\rightarrow Y\,$ such that
\qq
\pi\circ s_h=\psi|_h\,,\qquad\pi\circ s_c=\psi|_c\,,\qquad\pi\circ s_b=\psi_b\,.
\qqq
Similarly as in (\ref{Witeq}), we have
\qq
&&\ee^{\frac{\ii}{2}\int\limits_\CT\psi^*H}\,=\,\ee^{\frac{\ii}{2}\sum\limits_h(
\int_h\psi^*H+\int_{\zeta(h)}\psi^*H)}\,=\,\ee^{\ii\sum\limits_h
\int_h\psi^*H}\,=\,\ee^{\ii\sum\limits_h
\int_hs_h^*dB}\,=\,\ee^{\ii\sum\limits_{c\subset h}
\int_cs_h^*B\,+\,\ii\sum\limits_{\zeta(c)\subset h}\int_{\zeta(c)}s_h^*B}\cr\cr
&&=\,\ee^{\ii\sum\limits_{c\subset h}
\int_cs_h^*B\,+\,\ii\hspace{-0.1cm}\sum\limits_{c\subset\zeta(h)}
\int_c(s_h\circ\zeta)^*B}
\,=\,\ee^{\ii\sum\limits_{c\subset h}
\int_cs_c^*B\,+\,\ii\sum\limits_{c\subset h}\int_h(s_h^*B-s_c^*B)\,+\,\ii
\hspace{-0.1cm}\sum\limits_{c\subset\zeta(h)}\int_cs_c^*B
\,+\,\ii\hspace{-0.1cm}\sum\limits_{c\subset\zeta(h)}\int_c((s_h\circ\zeta)^*B
-s_c^*B)}\cr\cr
&&=\,\ee^{\ii\sum\limits_{c\subset h}
\int_cs_c^*B\,+\,\ii\hspace{-0.1cm}\sum\limits_{c\subset\zeta(h)}\int_cs_c^*B}\,
\ee^{\ii\sum\limits_{c\subset h}\int_h(s_c,s_h)^*F_\CL}\,\ee^{\ii\hspace{-0.1cm}
\sum\limits_{c\subset\zeta(h)}\int_c(s_c,s_h\circ\zeta)^*F_\CK}\,=\,
\ee^{\ii\sum\limits_{c\subset\Sigma}\int_cs_c^*B}\,\prod\limits_{c\subset h}hol_\CL
((s_c,s_h)|_{\partial c})\cr
&&\times\prod\limits_{c\subset\zeta(h)}hol_\CK
((s_c,s_h\circ\zeta)|_{\partial c})\,\cong\,\ee^{\ii\sum\limits_{c\subset\Sigma}
\int_cs_c^*B}\,\Big(\mathop{\otimes}\limits_{b\subset c\subset h}hol_\CL
(s_c|_b,s_h|_b)\Big)\otimes\Big(\mathop{\otimes'}\limits_{\zeta(b)\subset c
\subset h}hol_\CL(s_c|_{\zeta(b)},s_h|_{\zeta(b)}\Big)\cr
&&\times\,\Big(\mathop{\otimes}\limits_{b\subset c\subset\zeta(h)}
hol_\CK((s_c|_b,s_h\circ\zeta|_b)\Big)\otimes
\Big(\mathop{\otimes'}\limits_{\zeta(b)\subset c
\subset\zeta(h)}hol_\CK(s_c|_{\zeta(b)},s_h\circ\zeta|_{\zeta(b)})\Big)\,\cong\,
\ee^{\ii\sum\limits_{c\subset\Sigma}\int_cs_c^*B}\cr
&&\times\,\Big(\mathop{\otimes}\limits_{b\subset c\subset h}\big(hol_\CL
(s_c|_b,s_b)\otimes hol_\CL(s_b,s_h|_b)\big)\Big)\otimes\Big(\mathop{\otimes'}
\limits_{\zeta(b)\subset c\subset h}\big(hol_\CK(s_c\circ\zeta|_b,s_b)\otimes
hol_\CK(s_b,s_h\circ\zeta|_b)\big)\Big)\cr
&&\otimes\Big(\mathop{\otimes}\limits_{b\subset c\subset\zeta(h)}
\big(hol_\CL(s_c|_b,s_b)\otimes hol_\CK(s_b,s_h\circ\zeta|_b)\big)\Big)\otimes
\Big(\mathop{\otimes'}\limits_{\zeta(b)\subset c\subset\zeta(h)}
\big(hol_\CK(s_c\circ\zeta|_b,s_b)\otimes hol_\CL(s_b,s_h|_b)\big)\Big),\,\quad
\qqq
where $\,\otimes'\,$ means that $\,b\subset\CT'\,$ are omitted to avoid
an overcount. Reshuffling the terms on the right hand side, we obtain
\qq
&&\ee^{\frac{\ii}{2}\int\limits_\CT\psi^*H}\,\cong\,
\ee^{\ii\sum\limits_{c\subset\Sigma}\int_cs_c^*B}\,
\Big(\mathop{\otimes}\limits_{b\subset c\subset h}hol_\CL
(s_c|_b,s_b)\Big)\otimes\Big(\mathop{\otimes}\limits_{b\subset c\subset\zeta(h)}
\big(hol_\CL(s_c|_b,s_b)\Big)\cr
&&\hspace{3.2cm}\otimes\,\Big(\mathop{\otimes'}
\limits_{\zeta(b)\subset c\subset h}\big(hol_\CK(s_c\circ\zeta|_b,s_b)\Big)
\otimes\Big(\mathop{\otimes'}\limits_{\zeta(b)\subset c\subset\zeta(h)}
\big(hol_\CK(s_c\circ\zeta|_b,s_b)\Big)\cr
&&\hspace{3.2cm}\otimes\,\Big(\mathop{\otimes}\limits_{b\subset c\subset h}
hol_\CL(s_b,s_h|_b)\Big)\otimes\Big(\mathop{\otimes'}\limits_{b\subset\zeta(c)
\subset h}hol_\CL(s_b,s_h|_b)\Big)\cr
&&\hspace{3.2cm}\otimes\,\Big(\mathop{\otimes'}
\limits_{b\subset\zeta(c)\subset\zeta(h)}
hol_\CK(s_b,s_h\circ\zeta|_b)\Big)
\otimes\Big(\mathop{\otimes}\limits_{b\subset c
\subset\zeta(h)}hol_\CK(s_b,s_h\circ\zeta|_b)\Big)\cr\cr
&&\hspace{1.4cm}\cong\,\ee^{\ii\sum\limits_{c\subset\Sigma}\int_cs_c^*B}\,
\Big(\mathop{\otimes}\limits_{b\subset c\subset\Sigma}hol_\CL
(s_c|_b,s_b)\Big)\otimes\Big(\mathop{\otimes}
\limits_{\zeta(b)\subset c\subset\Sigma}\big(hol_\CK(s_c\circ\zeta|_b,s_b)\Big)\cr
&&\hspace{3.2cm}\otimes\,\Big(\mathop{\otimes}\limits_{\substack{b\subset c\subset h
\\ b\subset\CT'}}
hol_\CL(s_b,s_h|_b)\Big)\otimes
\Big(\mathop{\otimes}\limits_{\substack{b\subset\zeta(c)\subset h\\
b\subset\CT'}}hol_\CK(s_b,s_h|_b)\Big),
\label{hfWiteq}
\qqq
where we used the fact that, in the term between $\,\cong\,$ signs, 
in the first line
for each pair $\,(b,c)\,$ such that $\,b\subset c\not\in\Sigma\,$
and in the second line for $\,(b,c)\,$ such that $\,\zeta(b)\subset c\not
\in\Sigma\,$ there are two of tetrahedra $\,h\,$ or $\,\zeta(h)\,$
containing $\,c\,$ inducing on it opposite orientations.
Similarly in the third line for each $\,(b,h)\,$ such that
$\,b\not\subset\CT'\,$ and $\,b\subset h\,$ and in the forth line
for each $\,(b,h)\,$ such that $\,b\not\subset\CT'\,$ and
$\,b\subset\zeta(h)\,$ there are two of triangles $\,\zeta(c)\,$ or
$\,c\,$ in $\,h\,$ containing $\,b\,$ and inducing on it opposite
orientations. We still have to analyze the last line of (\ref{hfWiteq}).
To this end, let us consider a small $\,\zeta$-invariant neighborhood
$\,U\subset\CT\,$ around a fixed  edge $\,b\subset\CT'\,$ that
is diffeomorphic to $\,I\times D\,$ where $\,I\,$ is an interval
and $\,D\,$ is a unit disc in the complex plane, with $\,\zeta\,$
acting by $\,(t,z)\mapsto(t,-z)\,$ so that $\,\CT'\cap U\,$
is represented by $\,I\times\{0\}$. \,We may assume that the
lifts $\,h,\,c\,$ of tetrahedra $\,\tilde h\,$ and triangles $\,\tilde c\,$
that share the edge $\,\tilde b\,$ are chosen so that their intersections
with the disc $\,\{x\}\times D\,$ transverse to $\,b\,$ are as on
\,Fig.\,\ref{fig:aroundedge} (otherwise, we change those lifts). 
\noindent Then the
contribution of $\,b\,$ (considered with the orientation of the interval
$\,I$) \,to the last line of (\ref{hfWiteq}) is
\qq
&hol_\CL(s_b,s_{h_1}|_b)\otimes hol_\CL(s_{h_2}|_b,s_b)\otimes
hol_\CL(s_b,s_{h_2})\otimes\,\cdots\,\otimes hol_\CL(s_b,s_{h_k}|_b)
\otimes hol_\CK(s_{h_1},s_b)&\cr
&\mathop{\cong}\limits^{\tilde t}\,hol_\CK(s_b,s_b)\,=\,hol_{\CN'}(s_b)\,
\cong\,hol_{N'}(\psi|_b)&
\qqq
where we used the fact that $\,\psi(\CT')\subset M'\,$ due to the
equivariance (\ref{equivcond2}) so that $\,s_b\,$ takes values in $\,Y'$.
Hence the last line in (\ref{hfWiteq}) builds to
\qq
hol_{N'}(\psi|_{\CT'})\ \in\,\mathop{\otimes}\limits_{v\in\partial\CT'}
N'^{\pm1}_{\psi(v)}\,\cong\,\mathop{\otimes}\limits_{v\in\Sigma'}N'^{-1}_{\phi(v)}\,,
\label{2ndline}
\qqq
where the last isomorphism uses the fact that $\,\partial\CT'
=\CT'\cap\Sigma=\Sigma'\,$ and the natural isomorphism 
$\,N'\cong N'^{-1}$,
\,see (\ref{trivCN'2}) and the remark under (\ref{19}). 
In fact, the last line in (\ref{hfWiteq}) results in (\ref{2ndline})
for any choice of the lifts $\,h,\,c$.
Recall from (\ref{splitv}), (\ref{27}) and (\ref{CPN'}) 
that the first line
in (\ref{hfWiteq}) may be naturally interpreted as an element of the line
$\,\mathop{\otimes}_{v\in\Sigma'}N'_{\phi(v)}$, \,which is consistent 
with the fact
that the tensor product of both lines describes a $\,U(1)$-phase. \,Now,
if $\,M'\,$ is a 1-connected submanifold of $\,M\,$ then $\,N'\cong
\mathbb C\,$ as a flat line bundle and both lines of (\ref{hfWiteq}) may
be viewed as contributing $\,U(1)$-phases, the first one equal to
$\,\sqrt{\Hol_\CG(\phi)}\,$ and the second one equal to $\,1$. \,This proves
the identity (\ref{sqrtSWZ}).
\vskip -0.1cm

\hspace{14cm}$\blacksquare$
\vskip 0.3cm

\noindent{\bf Remark.} \,The right hand side of (\ref{sqrtSWZ}) could
be taken as Witten-type definition of the Feynman amplitude
$\,\ee^{\frac{\ii}{2} S_\WZ(\phi)}\,$ for equivariant maps $\,\phi$. \,The
net result of imposing the equivariance condition (\ref{equivcond2})
on the extension $\,\psi\,$ of $\,\phi\,$ is to make $\,\int_\CT\psi^*H\,$
well defined modulo $\,4\pi\,$ rather than $\,2\pi$.

\vskip 0.4cm

\nsection{Local data}
\label{sec:loc_data}

\noindent Let $\,L\,$ be a line bundle over $\,M\,$ (as always here,
equipped with a hermitian structure and a unitary connection)
and $\,(O_i)_{i\in J}\,$ a sufficiently fine open covering of
$\,M$. \,It is well known that $\,L\,$ may be represented (up to
an isomorphism) by local data $\,(a_i,h_{ij})\,$ with real connection
1-forms $\,a_i\,$ and $\,U(1)$-valued transition functions, defined 
on $\,O_i\,$ and $\,O_{i_1}\cap O_{i_2}\equiv O_{i_1i_2}$, \,respectively. 
\,Such local data are obtained from local normalized sections 
$\,\sigma_i:O_i\rightarrow L\,$ of $\,L\,$ by the formulae
\qq
\ii\nabla_{\hspace{-0.05cm}L}\sigma_i=a_i\sigma_i\,,\qquad \sigma_{i_2}=h_{i_1i_2}
\sigma_{i_1}\,.
\qqq 
They satisfy the relations
\qq
a_{i_2}-a_{i_1}=\ii\,d\ln{h_{i_1i_2}}\,,\qquad h_{i_1i_2}h_{i_2i_3}=h_{i_1i_3}\,.
\qqq
The curvature closed 2-form $\,F\,$ of $\,L\,$ is then equal to $\,da_i\,$ 
on $\,O_i$. \,If $\,L\,$ is trivializable, i.e. it possesses a global 
normalized flat section $\,\sigma:M\rightarrow L$, \,then 
$\,\sigma_i=l_i\sigma\,$ on $\,O_i\,$ with $\,(l_i)\,$ providing 
local data for the section $\,\sigma\,$ that satisfy the relations
\qq
a_i=\ii\,dl_i\,,\qquad h_{i_1i_2}=(l_{i_1})^{-1}l_{i_2}\,.
\qqq
\vskip 0.1cm

Similarly one may extract local data for a gerbe
$\,\CG=(Y,B,\CL,t)\,$ over $\,M\,$ [\onlinecite{Murray},\,\onlinecite{GR}].
\,Let $\,s_i:O_i\rightarrow Y\,$ be maps such that
$\,\pi(s_i(x))=x\,$ and let $\,\sigma_{i_1i_2}:O_{i_1i_2}\equiv
O_{i_1}\cap O_{i_2}\rightarrow\CL\,$ be such that $\,\sigma_{i_1i_2}(x)
\in\CL_{s_{i_1}(x),s_{i_2}(x)}$.
\,We demand that $\,|\sigma_{i_1i_2}(x)|=1\,$ and $\,\sigma_{i_2i_1}(x)
=\sigma_{i_1i_2}(x)^{-1}\,$ (i.e. is the dual element to $\,\sigma_{i_1i_2}(x)$).
\,The local data $\,(B_i,A_{i_1i_2},g_{i_1i_2i_3})\,$
for the gerbe $\,\CG\,$ are then defined by the relations
\qq
B_i=s_i^*B\,,\qquad
\ii\,\nabla_{\hspace{-0.05cm}\CL}\sigma_{i_1i_2}=\sigma_{i_1i_2}A_{i_1i_2}\,,\qquad
t\circ(\sigma_{i_1i_2}\otimes\sigma_{i_2i_3})=g_{i_1i_2i_3}\,\sigma_{i_1i_3}\,.
\label{locd}
\qqq
$B_i\,$ are real 2-forms on $\,O_i$, $\,A_{i_1i_2}=-A_{i_2i_1}\,$ are real
1-forms on $\,O_{i_1i_2}\,$
and $\,g_{i_1i_2i_3}=g_{i_2i_3i_1}=g_{i_3i_1i_2}=g_{i_2i_1i_3}^{-1}=g_{i_1i_3i_2}^{-1}
=g_{i_3i_2i_1}^{-1}\,$
are $\,U(1)$-valued functions on $\,O_{i_1i_2i_3}$. \,One has the relations
\qq
B_{i_2}-B_{i_1}=dA_{i_1i_2}\,,\qquad A_{i_1i_2}-A_{i_1i_3}+A_{i_2i_3}=\ii\,
d\ln{g_{i_1i_2i_3}}\,,\qquad
g_{i_1i_2i_3}\,g_{i_1i_2i_4}^{-1}\,g_{i_1i_3i_4}\,g_{i_2i_3i_4}^{-1}=1\,.
\label{delcoh}
\qqq
The curvature $\,H\,$ of $\,\CG\,$ satisfies $\,H=dB_i\,$ on $\,O_i$. 
\,In terms of the local data, the expression for the gerbe holonomy
takes the form \cite{GR}
\qq
\Hol_{\CG}(\phi)\,=\,\exp\Big[\sum\limits_c\int_c\phi^*B_{i_c}+\,
\ii\sum\limits_{b\subset c}\int_b\phi^*A_{i_ci_b}\Big]\prod\limits_{v\in b\subset c}
g_{i_ci_bi_v}(\phi(v))^{\pm1}\,,
\qqq
where $\,c$, $\,b\,$ and $\,v\,$ are the triangles, edges and vertices
of a sufficiently fine triangulation of $\,\Sigma\,$ and the indices
$\,i_c$, $\,i_b\,$ and $\,i_v\,$ are chosen so that
\qq
\phi(c)\subset O_{i_c}\,,\qquad\phi(b)\subset O_{i_b}\,,\qquad
\phi(v)\in O_{i_v}\,.
\label{i's}
\qqq
\vskip 0.1cm

Let now $\,\Theta\,$ be an involution of $\,M$. \,We shall assume that
the covering $\,(O_i)_{i\in J}\,$ is invariant under the $\,\Zb$-action
induced by the involution $\,\Theta$, \,i.e. $\,zO_i\,=\,O_{zi}\,$ for some
$\,\Zb$-action on the index set $\,J$. \,We shall write $\,-z$, $\,-x$,
$\,-i\,$ for $\,(-1)z$, $\,(-1)x=\Theta(x)\,$ and $\,(-1)i$.
\,Let $\,\tilde G=(\tilde Y,\tilde B,\tilde\CL,
\tilde t)\,$ be a $\,\Zb$-equivariant extension of $\,\CG$. \,We shall repeat
the previous construction of local data for $\,\tilde\CG\,$ defining
maps $\,\tilde s_i^z:O_i\rightarrow\tilde Y\,$ such that
$\,\tilde s_i^z(x)=(z,s_{z^{-1}i}(z^{-1}x))\,$ (of course $z^{-1}=z\,$ in 
$\,\mathbb Z_2$) \,and $\,\tilde\sigma_{i_1i_2}^{z_1z_2}:O_{i_1i_2}
\rightarrow\tilde\CL\,$ such that
$\,\tilde\sigma_{i_1i_2}^{z_1z_2}(x)\in\tilde\CL_{\tilde s_{i_1}^{z_1}(x),\tilde s_{i_2}^{z_2}(x)}$,
$\,|\tilde\sigma_{i_1i_2}^{z_1z_2}(x)|=1\,$ and $\,\tilde\sigma_{i_2i_1}^{z_2z_1}(x)
=\tilde\sigma_{i_1i_2}^{z_1z_2}(x)^{-1}$. Then the local data
$\,(\tilde B_i^z,\tilde A_{i_1i_2}^{z_1z_2},\tilde g_{i_1i_2i_3}^{z_1z_2z_3})\,$ are defined 
by the identities similar to (\ref{locd}):
\qq
\tilde B_i^z=(\tilde s_i^z)^*\tilde B\,,\qquad
\ii\nabla_{\hspace{-0.05cm}\tilde\CL}\tilde\sigma_{i_1i_2}^{z_1z_2}=\tilde\sigma_{i_1i_2}^{z_1z_2}
\tilde A_{i_1i_2}^{z_1z_2}\,,\qquad
\tilde t\circ(\tilde\sigma_{i_1i_2}^{z_1z_2}\otimes\tilde\sigma_{i_2i_3}^{z_2z_3})=
\tilde g_{i_1i_2i_3}^{z_1z_2z_3}\,\tilde\sigma_{i_1i_3}^{z_1z_3}
\label{fi0}
\qqq
and they satisfy relations similar to (\ref{delcoh}):
\qq
\tilde B_{i_2}^{z_2}-B_{i_1}^{z_1}=d\tilde A_{i_1i_2}^{z_1z_2},
\quad\tilde A_{i_1i_2}^{z_1z_2}-\tilde A_{i_1i_3}^{z_1z_3}+\tilde A_{i_2i_3}^{z_2z_3}=\ii\,
d\ln{\tilde g_{i_1i_2i_3}^{z_1z_2z_3}},\quad
\tilde g_{i_1i_2i_3}^{z_1z_2z_3}(\tilde g_{i_1i_2i_4}^{z_1z_2z_4})^{-1}
\tilde g_{i_1i_3i_4}^{z_1z_3z_4}(\tilde g_{i_2i_3i_4}^{z_2z_3z_4})^{-1}\hspace{-0.05cm}=1.
\quad\ \label{hatdelcoh}
\qqq
One has:
\qq
\tilde B_i^1=B_i\,,\qquad\tilde A_{i_1i_2}^{1\,1}=A_{i_1i_2}\,,\qquad
\tilde g_{i_1i_2i_3}^{1\,1\,1}=g_{i_1i_2i_3}\,.
\qqq
\vskip 0.1cm

The local data $\,(\tilde B_i^z,\tilde A_{i_1i_2}^{z_1z_2},\tilde g_{i_1i_2i_3}^{z_1z_2z_3})\,$
may be reduced to simpler ones that provide local data for a 
$\,\mathbb Z_2$-equivariant structure on gerbe $\,\CG\,$ as defined 
in [\onlinecite{GSW},\,\onlinecite{GenWar}]. To this end, we shall impose
the identities
\qq
&&\tilde\sigma_{i_1i_2}^{(-1)(-1)}\,=\,(-1)\tilde\sigma_{(-i_1)(-i_2)}^{1\,1}\circ\Theta\,,
\label{1stch}\\
&&\tilde\sigma_{i_1i_2}^{1(-1)}=\tilde t\circ(\tilde\sigma_{i_1i_2}^{1\,1}
\otimes\tilde\sigma_{i_2i_2}^{1(-1)})\qquad{\rm i.e.}\qquad\tilde g_{i_1i_2i_2}^{1\,1(-1)}=1
\label{2ndch}
\qqq
and shall define 1-forms $\,\Pi_i\,$ on $\,O_i$, $\,U(1)$-valued functions
$\,\chi_{i_ii_2}=\chi_{i_2i_1}^{-1}\,$ on $\,O_{i_1i_2}\,$ and $\,f_i\,$
on $\,O_i\,$ by the relations
\qq
\Pi_i=\tilde A_{i\,i}^{1(-1)}\,,\qquad
\chi_{i_1i_2}=\tilde g_{i_1i_1i_2}^{(-1)1(-1)}\,,\qquad
(-1)\tilde\sigma_{i\,i}^{1(-1)}=f_i\,\tilde\sigma_{(-i)(-i)}^{(-1)1}\circ\Theta\,.
\label{deffi}
\qqq
Let us notice that from the last of the relations (\ref{fi0})
it follows that
\qq
1\,=\,\tilde g_{i_1i_2i_2}^{(-1)(-1)1}\,(\tilde g_{i_1i_2i_2}^{1(-1)1})^{-1}\,
\tilde g_{i_1i_1i_2}^{1(-1)1}\,(\tilde g_{i_1i_1i_2}^{1(-1)(-1)})^{-1}\,=\,
\chi_{i_2i_1}\,\chi_{i_1i_2}
\qqq
i.e. $\,\chi_{i_2i_1}=\chi_{i_1i_2}^{-1}$.
\vskip 0.4cm

\noindent{\bf Proposition 5.} \ We have the following identities:
\qq
&&\Theta^*B_{-i}=B_i+d\Pi_i\,,\label{BPi}\\
&&\Theta^*A_{(-i_1)(-i_2)}=A_{i_1i_2}+\Pi_{i_2}-\Pi_{i_1}-\ii\,d\ln{\chi_{i_1i_2}}\,,
\label{APichi}\\
&&\Theta^*g_{(-i_1)(-i_2)(-i_3)}=g_{i_1i_2i_3}\,(\chi_{i_1i_2})^{-1}\,\chi_{i_1i_3}\,
(\chi_{i_2i_3})^{-1}\,,\label{gchi}\\
&&\Theta^*\Pi_{-i}+\Pi_i=\ii\,d\ln{f_i}\,,\label{-PiPi}\\
&&\Theta^*\chi_{(-i_1)(-i_2)}\,\chi_{i_1i_2}=(f_{i_1})^{-1}\,f_{i_2}\,,\label{-chichi}\\
&&\Theta^*f_{-i}=f_i\,.\label{-fifi}
\qqq
\vskip 0.3cm

\noindent{\bf Proof.} \ From the first of the relations (\ref{fi0}),
we have: 
\qq
\tilde B_i^{(-1)}=\Theta^*B_{-i}\,.
\label{fi1}
\qqq
On the other hand, (\ref{hatdelcoh}) and the first of the definitions
(\ref{deffi}) imply
that
\qq
\tilde B_i^{(-1)}-\tilde B_i^1=d\tilde A_{i\,i}^{1(-1)}=d\Pi_i
\label{fi2}
\qqq
so that (\ref{BPi}) follows.

Next, since the action of $\,-1\,$ on $\,\tilde\CL\,$ preserves the connection,
we infer from (\ref{1stch}) and the $2^{\rm nd}$ of the definitions
(\ref{fi0}) that
\qq
\tilde A_{i_ii_2}^{(-1)(-1)}=\Theta^*A_{(-i_1)(-i_2)}\,.
\qqq
Now, the $2^{\rm nd}$ of the relations (\ref{hatdelcoh}) gives:
\qq
\tilde A_{i_1\,i_1}^{1(-1)}-\tilde A_{i_1\,i_2}^{1(-1)}+\tilde A_{i_1\,i_2}^{(-1)(-1)}=
\ii\,d\ln{\tilde g_{i_1i_1i_2}^{1(-1)(-1)}}\,,\ \quad
{\rm i.e.}\quad
\Theta^*A_{(-i_1)(-i_2)}=\tilde A_{i_1\,i_2}^{1(-1)}-\Pi_{i_1}-\ii\,d\ln{\chi_{i_1i_2}}\,.
\quad\label{fi3}
\qqq
Similarly, \,by (\ref{2ndch}),
\qq
\tilde A_{i_1i_2}^{1\,1}-\tilde A_{i_1i_2}^{1(-1)}+\tilde A_{i_2i_2}^{1(-1)}=\ii\,
d\ln{\tilde g_{i_1\,i_2\,i_2}^{1\,1(-1)}}=0\,,\quad\ {\rm i.e.}\quad
\tilde A^{1(-1)}_{i_1i_2}=A_{i_1i_2}+\Pi_{i_2}\,.
\label{fi4}
\qqq
Substituted to (\ref{fi3}), this implies (\ref{APichi}).
\vskip 0.1cm

From the last of the relation (\ref{1stch}), the commutation of the
$\,\Zb\,$ action with the groupoid multiplication $\,\tilde t\,$ and
the definition (\ref{fi0}), it follows that
\qq
\tilde g_{i_1\,i_2\,i_3}^{(-1)(-1)(-1)}=\Theta^*g_{(-i_1)(-i_2)(-i_3)}
\qqq
and from the last of the relations (\ref{hatdelcoh}) that
\qq
&&\tilde g_{i_1\,i_2\,i_3}^{(-1)(-1)(-1)}=\tilde g_{i_1\,i_2\,i_3}^{1(-1)(-1)}\,
(\tilde g_{i_1\,i_1\,i_3}^{1(-1)(-1)})^{-1}\,\tilde g_{i_1\,i_i\,i_2}^{1(-1)(-1)}\,=\,
\tilde g_{i_1\,i_2\,i_3}^{1(-1)(-1)}\,\chi_{i_1i_3}\,(\chi_{i_1i_2})^{-1}\,,\cr
&&\tilde g_{i_1i_2i_3}^{1(-1)(-1)}\,=\,\tilde g_{i_1i_2i_3}^{1\,1(-1)}\,
(\tilde g_{i_1i_2i_2}^{1\,1(-1)})^{-1}\,\tilde g_{i_2i_2i_3}^{1(-1)(-1)}\,=\,
\tilde g_{i_1i_2i_3}^{1\,1(-1)}\,(\chi_{i_2i_3})^{-1}\,,\cr
&&\tilde g_{i_1i_2i_3}^{1\,1(-1)}\,=\,g_{i_1i_2i_3}\,\tilde g_{i_1i_3i_3}^{1\,1(-1)}\,
(\tilde g_{i_2i_3i_3}^{1\,1(-1)})^{-1}\,=\,g_{i_1i_2i_3}\,.
\qqq
The latter four relations imply (\ref{gchi}).
\vskip 0.1cm

From the middle one of the definitions (\ref{fi0}) and the last one of
(\ref{deffi}), it follows that
\qq
&&-\Theta^*\tilde A_{(-i)(-i)}^{1(-1)}\,=\,\Theta^*\tilde A_{(-i)(-i)}^{(-1)1}
\,=\,A_{i\,i}^{1(-1)}\,-\,\ii\,d\ln{f_{i_1}}\,.
\qqq
This gives (\ref{-PiPi}).
\vskip 0.1cm

In order to prove (\ref{-chichi}) and (\ref{-fifi}), let us additionally 
define $\,U(1)$-valued functions
$\,f_{i_1i_2}\,$ on $\,O_{i_1i_2}\,$ by the relation
\qq
(-1)\tilde\sigma_{i_1i_2}^{1(-1)}=f_{i_1i_2}\,
\tilde\sigma_{(-i_1)(-i_2)}^{(-1)(1)}\circ\Theta
\label{fi1i2}
\qqq
so that, in particular, $\,f_{i\,i}=f_i$. \,Applying $\,-1\,$ to both sides
of (\ref{fi1i2}) and composing the result with the action of $\,\Theta$, 
\,we obtain the relation
\qq
\tilde\sigma_{i_1i_2}^{1(-1)}\circ\Theta\,=\,\Theta^*f_{i_1i_2}\,(-1)
\tilde\sigma_{(-i_1)(-i_2)}^{(-1)1}\,.
\qqq
The symmetry $\,\tilde\sigma_{i_2i_1}^{(-1)1}=(\tilde\sigma_{i_1i_2}^{1(-1)})^{-1}\,$
that is preserved by the action of $\,-1\,$ implies then that
\qq
\tilde\sigma_{i_2i_1}^{(-1)1}\circ\Theta\,=\,(\Theta^*f_{i_1i_2})^{-1}\,(-1)
\tilde\sigma_{(-i_2)(-i_1)}^{(-1)1}
\qqq
and, by comparison to (\ref{fi1i2}), that
\qq
\Theta^*f_{(-i_2)(-i_1)}=f_{i_1i_2}\,.
\label{-1fi1i2}
\qqq
In particular, setting $\,i_1=i_2=i$, \,we obtain (\ref{-fifi}).
\,Next, using the commutation of the action of $\,-1\,$ with $\,\tilde t$,
\,we obtain the identity
\qq
&&\tilde t\circ\big((-1)\tilde\sigma_{i_1i_2}^{1\,1}\otimes 
(-1)\tilde\sigma_{i_2i_2}^{1(-1)})\big)
=\tilde t\circ\big(\tilde\sigma_{(-i_1)(-i_2)}^{(-1)(-1)}\circ\Theta
\otimes f_{i_2}\,\tilde\sigma_{(-i_2)(-i_2)}^{(-1)1}\circ\Theta\big)\cr
&&=f_{i_2}\,\Theta^*\tilde g_{(-i_1)(-i_2)(-i_2)}^{(-1)(-1)1}\,
\tilde\sigma_{(-i_1)(-i_2)}^{(-1)1}\circ\Theta\cr
&&=(-1)\tilde t\circ\big(\tilde\sigma_{i_1i_2}^{1\,1}\otimes
\tilde\sigma_{i_2i_2}^{1(-1)})\big)=
\tilde g_{i_1i_2i_2}^{11(-1)}\,(-1)\tilde\sigma_{i_1i_2}^{1(-1)}=f_{i_1i_2}\,
\tilde\sigma_{(-i_1)(-i_2)}^{(-1)1}\circ\Theta\,.
\qqq
It follows that
\qq
f_{i_1i_2}\,=\,f_{i_2}\,\Theta^*\tilde g_{(-i_1)(-i_2)(-i_2)}^{(-1)(-1)1}\,=
\,f_{i_2}\,\Theta^*\chi_{(-i_2)(-i_1)}\,.
\qqq
Combining this with (\ref{-1fi1i2}), \,we infer that
\qq
f_{i_2}\,\Theta^*\chi_{(-i_2)(-i_1)}\,=\,\Theta^*f_{-i_1}\,\chi_{i_1i_2}
\qqq
that, together with (\ref{-fifi}), implies (\ref{-chichi}).
\vskip -0.1cm

\hspace{14cm}$\blacksquare$
\vskip 0.3cm

Conversely, we may recover
the local data $\,(\tilde B_i^z,\tilde A_{i_1i_2}^{z_1z_2},
\tilde g_{i_1i_2i_3}^{z_1z_2z_3})\,$ of a $\,\Zb$-equivariant extension 
$\,\tilde\CG\,$ of gerbe $\,\CG$ from the local data 
$\,(B_i,A_{i_1i_2},g_{i_1i_2i_3})\,$ and
$\,(\Pi_i,\chi_{i_1i_2},f_i)\,$ by setting
\qq
&&\tilde B_{i}^1=B_i\,,\qquad\tilde B_{i}^{-1}=\Theta^*B_{-i}\,,\label{imposeBs}\\
&&\tilde A_{i_1i_2}^{1\,1}=A_{i_1i_2}\,,\qquad\tilde A_{i_1i_2}^{1(-1)}=
A_{i_1i_2}+\Pi_{i_2}\,,\qquad\tilde A_{i_1i_2}^{(-1)(-1)}=
\Theta^*A_{(-i_1)(-i_2)}\,,\quad\label{imposeAs}\\
&&\tilde g^{1\,1\,1}_{i_1i_2i_3}=g_{i_1i_2i_3}\,,\qquad\tilde g^{1\,1(-1)}_{i_1i_2i_3}
=g_{i_1i_2i_3}\,,\qquad 
\tilde g^{1(-1)(-1)}_{i_1\,i_2\,i_3}=g_{i_1i_2i_3}\,(\chi_{i_2i_3})^{-1}\,,\cr
&&\tilde g_{i_1\,i_2\,i_3}^{(-1)(-1)(-1)}=\Theta^*g_{(-i_1)(-i_2)(-i_3)}\quad
\label{imposegs}
\qqq
and imposing the desired symmetry in the indices. The identities
(\ref{hatdelcoh}) follow then from the relations (\ref{BPi}), 
(\ref{APichi}) and (\ref{gchi}). 
\vskip 0.3cm

We shall also need to find  local data for the flat line
bundle $\,N'\,$ over the fixed-point manifold $\,M'\,$ of $\,\Theta$.
\,Let us denote $\,O_i\cap M'\equiv O'_i=O'_{-i}$. \,For
$\,x'\in O'_i$, $\,(s_i(x'),s_i(x'))\in Z$, \,see (\ref{Z}), and
$\,\tilde\sigma_{i(-i)}^{1(-1)}(x')\in\tilde\CL_{(1,s_i(x')),(-1,s_{i}(x'))}
=\CN'_{s_i(x')}$. \,We have
\qq
\ii\nabla_{\hspace{-0.05cm}\CN'}\tilde\sigma_{i(-i)}^{1(-1)}(x')\,=\,
\tilde A_{i(-i)}^{1(-1)}(x')\,\tilde\sigma_{i(-i)}^{1(-1)}(x')\,=\,
(A_{i(-i)}+\Pi_{-i})(x')\,\tilde\sigma_{i(-i)}^{1(-1)}(x')\,,
\qqq
and for $\,x'\in O'_{i_1i_2}$, 
\qq
\tilde t\big(\tilde\sigma_{i_1i_2}^{1\,1}(x')\otimes\tilde\sigma_{i_2(-i_2)}^{1(-1)}(x')\big)
\,=\,\tilde g_{i_1i_2(-i_2)}^{1\,1(-1)}(x')\,\tilde\sigma_{i_1(-i_2)}^{1(-1)}(x')\,.
\qqq
On the other hand,
\qq
\tilde t\big(\tilde\sigma_{i_1(-i_1)}^{1(-1)}(x')\otimes
\tilde\sigma_{(-i_1)(-i_2)}^{(-1)(-1)}(x'))\,=\,\tilde g_{i_1(-i_1)(-i_2)}^{1(-1)(-1)}(x')\,
\tilde\sigma_{i_1(-i_2)}^{1(-1)}(x')\,.
\qqq
From (\ref{nu'}), (\ref{nu'1}) and (\ref{1stch}), we infer that
\qq
\nu'\big(\tilde\sigma_{i_1i_2}^{1\,1}(x')\otimes
\tilde\sigma_{i_2(-i_2)}^{1(-1)}(x')\big)\,=\,
\frac{\tilde g_{i_1i_2(-i_2)}^{1\,1(-1)}(x')}{\tilde g_{i_1(-i_1)(-i_2)}^{1(-1)(-1)}(x')}
\,\tilde\sigma_{i_1(-i_1)}^{1(-1)}(x')\otimes
\tilde\sigma_{i_1i_2}^{1\,1}(x')\ \cr
=\,\frac{g_{i_1i_2(-i_2)}(x')\,
\chi_{(-i_1)(-i_2)}(x')}{g_{i_1(-i_1)(-i_2)}(x')}\,
\tilde\sigma_{i_1(-i_1)}^{1(-1)}(x')\otimes\tilde\sigma_{i_1i_2}^{1\,1}(x')\,.
\qqq
It follows that, as elements of the line bundle $\,N'\,$ over $\,M'$,
\qq
\tilde\sigma_{i_2(-i_2)}^{1(-1)}(x')\,=\,\frac{g_{i_1i_2(-i_2)}(x')
\,\chi_{(-i_1)(-i_2)}(x')}{g_{i_2(-i_2)(-i_2)}(x')}\,
\,\tilde\sigma_{i_1(-i_1)}^{1(-1)}(x')\,.
\qqq
Hence $\,(a'_i,h'_{i_1i_2})$, \,where
\qq
a'_i\,=\,\big(A_{i(-i)}+\Pi_{-i}\big)|_{O'_i}\,,\qquad
h'_{i_1i_2}\,=\,\frac{g_{i_1i_2(-i_2)}\,
\chi_{(-i_1)(-i_2)}}{g_{i_1(-i_1)(-i_2)}}\Big|_{O'_{i_1i_2}}\,,
\qqq
are local data of the flat line bundle $\,N'$. \,They satisfy
the relations
\qq
da'_i=0\,,\qquad a'_{i_2}-a'_{i_1}=\ii\,d\ln{h'_{i_1i_2}}\,,\qquad
h'_{i_1i_2}\,h'_{i_2i_3}=h'_{i_1i_3}
\qqq
that are straightforward to check.  
\,Let us define
\qq
l'_i=\big(\chi_{i(-i)}f_i\big)|_{O'_i}\,.
\qqq
Then
\qq
&&\ii\,d\ln{l'_i}=\big(\ii\,d\ln{\chi_{i(-i)}}+\ii\,d\ln{f_{-i}}\big)|_{O'_i}
=\big(A_{i(-i)}-\Theta^*A_{(-i)i}+\Pi_{-i}-\Pi_i+\Theta^*\Pi_{-i}+\Pi_i\big)|_{O'_i}\cr
&&\hspace{5.62cm}=2\big(A_{i(-i)}+\Pi_{-i})|_{O'_i}=2a'_i\,,
\qqq
and
\qq
&&(l'_{i_1})^{-1}l'_{i_2}=\big(\chi_{i_1(-i_1)})^{-1}(f_{i_1})^{-1}\chi_{i_2(-i_2)}f_{i_2}
\big)|_{O'_{i_1i_2}}\cr
&&=\big(\chi_{i_1(-i_1)})^{-1}\chi_{i_1(-i_2)}(\chi_{(-i_1)(-i_2)})^{-1}
(f_{i_1})^{-1}\chi_{i_2(-i_2)}\chi_{i_1(-i_2)}^{-1}\chi_{(-i_1)(-i_2)}f_{i_2}
\big)|_{O'_{i_1i_2}}\cr
&&=\big(\Theta^*g_{(-i_1)i_1i_2}\,(g_{i_1(-i_1)(-i_2)})^{-1}
\chi_{i_2(-i_2)}\chi_{i_1(-i_2)}^{-1}\chi_{(-i_1)(-i_2)}
\Theta^*\chi_{(-i_1)(-i_2})\chi_{i_1i_2}\big)|_{O'_{i_1i_2}}\cr
&&=\big(g_{(-i_1)i_1i_2}\,(g_{i_1(-i_1)(-i_2)})^{-1}
(\Theta^*g_{(-i_1)(-i_2)i_2})^{-1}g_{i_1i_2(-i_2)}(\chi_{(-i_1)(-i_2)})^2\big)|_{O'_{i_1i_2}}\cr
&&=\big(g_{(-i_1)i_1i_2}\,(g_{i_1(-i_1)(-i_2)})^{-1}
(g_{(-i_1)(-i_2)i_2})^{-1}g_{i_1i_2(-i_2)}(\chi_{(-i_1)(-i_2)})^2\big)|_{O'_{i_1i_2}}\cr
&&=\big((g_{i_1(-i_1)(-i_2)})^{-2}
(g_{i_1i_2(-i_2)})^2(\chi_{(-i_1)(-i_2)})^2\big)|_{O'_{i_1i_2}}=(h'_{i_1i_2})^2\,.
\qqq
The family $\,(l'_i)\,$ provides local data for the trivialization
of the flat line bundle $\,(N')^2\,$ considered in 
Sec.\,\ref{sec:sqrt_hol}. \,If $\,M'\,$ is 1-connected then
one may choose the square roots $\,\sqrt{l'_i}\,$ so that
\qq
\displaystyle{\ii\,d\ln{\sqrt{l'_i}}=a'_i\,,
\qquad (\sqrt{l'_i}_{\hspace{-0.04cm}_{1}})^{-1}\sqrt{l'_i}_{\hspace{-0.03cm}_{2}}
=h'_{i_1i_2}\,.}
\qqq
Such $\,(\sqrt{l'_i})$, \,defined modulo a global sign, \,provide local 
data for a trivialization of the flat line bundle $\,N'\,$ used
in Sec.\,\ref{sec:sqrt_hol}.
\vskip 0.1cm

We have now all elements at hand to present the local-data formula
for the square of the gerbe holonomy defined in 
Sec.\,\ref{sec:sqrt_hol}. A straightforward but somewhat
tedious verification that we omit here shows that
\qq
&&\sqrt{\Hol_\CG(\phi)}\cr\cr
&&=\,\exp\Big[\ii\sum\limits_c\int_c\phi^*B_{i_c}\,
+\,\ii\sum\limits_{b\subset c}\int_b\phi^*A_{i_ci_b}
+\,\ii\sum\limits_{b\subset\vartheta(c)}
\int_b\phi^*A_{(-i_c)i_b}\,-\,\ii\sum\limits_{b\subset\vartheta(c)}\int_b
\phi^*\Pi_{-i_c}\Big]\cr
&&\times\,\prod\limits_{v\in b\subset c}(g_{i_ci_bi_v}(\phi(v)))^{\pm1}
\prod\limits_{v\in b\subset\vartheta(c)}
\big((g_{(-i_c)i_bi_v}(\phi(v)))^{\pm1}
(\chi_{(-i_c)i_v}(\phi(v)))^{\pm1}\big)\mathop{{\prod}'}
\limits_{\substack{v\in b\subset\vartheta(c)\\ v\not\in\Sigma'}}
(f_{i_v}(\phi(v)))^{\mp1}\cr
&&\times\,\hspace{-0.19cm}
\prod\limits_{\substack{v\in b\subset\vartheta(c)\\ v\in\Sigma'}}
\big((g_{i_ci_v(-i_v)}(\phi(v)))^{\mp1}\,(\chi_{i_v(-i_v)}(\phi(v)))^{\pm1}
\,(\sqrt{l'_i}_{\hspace{-0.03cm}_{v}}(\phi(v)))^{\mp1}\big),
\qqq
where $\,c\,$ and $\,b\,$ run, as before, over the selected lifts
of triangles $\,\tilde c\,$ and $\,\tilde b\,$ of a sufficiently fine
orbifold triangulation of $\,\tilde\Sigma$, \,whereas $\,v\,$ runs
over all vertices projecting to vertices $\,\tilde v\,$ with 
the exception of the product $\,{\prod}'\,$ where only one of 
two vertices $\,v\not\in\Sigma'\,$ projecting to each vertex
$\,\tilde v\,$ is taken into consideration. The indices $\,i_c$, 
$\,i_b,$ and $\,i_v\,$ are chosen so that the relations (\ref{i's}) 
hold and, that, additionally, $\,i_{\vartheta(v)}=-i_v\,$ for 
$\,v\not\in\Sigma'$.

\nsection{$3d$-index}
\label{sec:3d_index}

\noindent Let, as above, $\,M\,$ be a manifold equipped with an involution
$\,\Theta\,$ whose fixed-point set $\,M'\,$ is a 1-connected submanifold
of $\,M$. Let $\,\CG=(Y,B,\CL,t)\,$ be a gerbe on $\,M\,$ 
with curvature $\,H=\Theta^*H\,$ and let $\,\tilde\CG=(\tilde Y,\tilde B,
\tilde\CL,\tilde t)\,$ be a $\,\mathbb Z_2$-equivariant extension of 
$\,\CG$. 
\vskip 0.1cm

Let $\,\CR\,$ be a compact oriented $\,3d$-manifold with an 
orientation-reversing involution $\,\rho\,$ with the derivative equal
to $\,-I\,$ at the fixed points. The fixed-point set $\,\CR'\,$ of
$\,\rho\,$ must then be discrete. As the main example of such a 
$\,3d$-manifold with involution, we shall consider the $3d$-torus 
$\,\mathbb T^3=\mathbb R^3/(2\pi\mathbb Z^3)\,$, viewed as the cube 
$\,[-\pi,\pi]^3\,$ with the periodic identifications, \,with $\,\rho\,$ 
given by $\,k\mapsto-k$.
\vskip 0.1cm

Let $\,\Phi:\CR\rightarrow M\,$ be a map satisfying the equivariance condition
\qq
\Phi\circ\rho\,=\,\Theta\circ\Phi\,.
\label{equivPhi}
\qqq   
We shall again view $\,\CR/\mathbb Z_2=\tilde\CR\,$ as a $\,\mathbb Z_2\,$
orbifold, choosing for it a sufficiently fine orbifold triangulation,
with simplices $\,\tilde h,\,\tilde c,\,\tilde b,\,\tilde v$, \,that lifts
to a triangulation of $\,\CR\,$ with simplices $\,h,\,c,\,b,\,v$, \,the ones
of positive dimension interchanged in pairs
by $\,\rho$. \,We shall select one $\,h$, one $\,c$ and one $\,b\,$ in
each such pair, together with the maps $\,s_h:h\rightarrow Y$,
$\,s_c:c\rightarrow Y\,$ and $\,s_b:b\rightarrow Y\,$ such that
\qq
\pi\circ s_h=\Phi|_h\,,\qquad\pi\circ s_c=\Phi|_c\,,
\qquad \pi\circ s_b=\Phi|_b\,.
\qqq
Given such choices, we shall consider the expression
\qq
&&\hspace{-0.6cm}K_\CG(\Phi)\,=\,
\ee^{-\frac{\ii}{2}\sum\limits_h\int_h\Psi^*H\,+\,\ii\sum\limits_{c\subset h}
\int_cs_c^*B}\Big(\mathop{\otimes}\limits_{b\subset c\subset h}
hol_\CL(s_c|_b,s_b)\Big)\otimes\Big(\mathop{\otimes}
\limits_{\rho(b)\subset c\subset h}hol_\CK(s_c\circ\rho|_b,s_b)\Big),
\label{CKPhi}
\qqq 
where only the selected $\,h,\,c\,$ and $\,b\,$ are considered,
the tetrahedra $\,h\,$ are taken with the orientation of $\,\CR\,$,
and $\,c\subset h$, $\,b\subset c\subset h$, and $\,\rho(b)\subset c
\subset h\,$ with the inherited orientations and, in the last case,
$\,b\,$ with the orientation related to that of $\,\rho(b)\,$ by $\,\rho$.
\,The contributions to the right hand side of (\ref{CKPhi}) from
triangles $\,c\,$ that are shared by two tetrahedra $\,h\,$ cancel out
as such $\,c\,$ appear twice with opposite orientations. Denote by
$\,\CF\,$ the simplicial complex that is the union of all selected
tetrahedra $\,h\,$ and by $\,\Sigma\,$ its boundary $\,\partial\CF\,$
which is the union of all triangles of the triangulation of $\,\CR\,$
that belong to only one of the selected $\,h$. \,If $\,c\subset\Sigma\,$
then also $\,\rho(c)\subset\Sigma\,$ so that $\,\Sigma\,$ is
$\,\rho$-invariant and $\,\rho\,$ preserves its orientation. \,The right
hand side of (\ref{CKPhi}) takes values in the line
\qq
\Big(\mathop{\otimes}\limits_{v\in b\subset c\subset\Sigma}
\CL_{s_c(v),s_b(v)}^{\pm 1}\Big)
\otimes\Big(\mathop{\otimes}\limits_{v\in\vartheta(b)
\subset c\subset\Sigma}\CK_{s_c(v),s_b(\vartheta(v))}^{\pm 1}\Big).
\label{lineK}
\qqq
If the simplicial complex $\,\CF\,$ forms a submanifold with boundary
of $\,\CR\,$ then $\,\Sigma=\partial\CF\,$ is a closed oriented surface
and the line (\ref{lineK}) is canonically isomorphic to $\,\mathbb C$,
\,as was shown in Sec.\,\ref{sec:sqrt_hol}. We may then identify
\qq
K_\CG(\Phi)\,=\,\ee^{-\frac{\ii}{2}\int_\CF\Phi^*H}\sqrt{\Hol_\CG(\phi|_{\partial\CF})}
\,\in\,U(1)\,.
\label{K(Phi)}
\qqq
For simplicity, we shall limit ourselves to such situations of which
an example is provided by $\,\CR=\mathbb T^3\,$ where we may take as
$\,\CF\,$ the subset obtained by restricting one of the components of
$\,k\in[-\pi,\pi]^3\,$ to non-negative (or non-positive)
values. Note that the right hand side of (\ref{K(Phi)}) squares
to $\,1\,$ by Lemma 1 of Sec.\,\ref{sec:sqrt_hol} and Proposition
2 of \,Sec.\,\ref{sec:gerbe_hol}. Hence $\,K_\CG(\phi)=\pm1$.
\vskip 0.4cm

\noindent{\bf Proposition 6.} \,The phase associated to
the right hand side (\ref{CKPhi}) is independent of the choice
of the simplices $\,h,\,c,\,b$, \,the maps $\,s_c,\,s_b\,$ and the orbifold
triangulation of $\,\tilde\CR$.
\vskip 0.3cm

\noindent{\bf Proof.} \,We shall proceed similarly as in the proof
of Proposition 3 in Sec.\,\ref{sec:sqrt_hol}, showing that the changes
of the selected simplices and/or of their maps to $\,Y\,$ lead
to expressions that are equivalent under the use of line-bundle
isomorphisms $\,\tilde t\,$ and the ones induced by the $\,\mathbb Z_2$
action, part of the structure of the gerbe $\,\tilde\CG$.
\vskip 0.1cm

1. \,If the tetrahedron $\,h_0\,$ is changed to $\,h'_0=\rho(h_0)\,$
then we have:
\qq
&&\ee^{-\frac{\ii}{2}\int_{h'_0}\Phi^*H}=\ee^{\frac{\ii}{2}\int_{h_0}\Phi^*H}=
\ee^{-\frac{\ii}{2}\int_{h_0}\Phi^*H}\,\ee^{\ii\int_{h_0}\Phi^*H}
=\ee^{-\frac{\ii}{2}\int_{h_0}\Phi^*H}
\,\ee^{\ii\int_{h_0}s_{h_0}^*dB}\cr\cr
&&=\ee^{-\frac{\ii}{2}\int_{h_0}\Phi^*H}\,
\ee^{\ii\sum\limits_{c\subset h_0}\int_cs_{h_0}^*B\,
+\,\ii\sum\limits_{\rho(c)\subset h_0}\int_{\rho(c)}
s_{h_0}^*B}=\ee^{-\frac{\ii}{2}\int_{h_0}\Phi^*H}\,\ee^{\ii\sum\limits_{c\subset h_0}
\int_cs_c^*B\,+\,\ii\sum\limits_{c\subset h_0}\int_c(s_c,s_{h_0})^*F_\CL}\cr\cr
&&\times\,\ee^{\ii\sum\limits_{\rho(c)\subset h_0}\int_cs_c^*B\,
+\,\ii\sum\limits_{\rho(c)\subset h_0}\int\limits_c(s_c,s_{h_0}\circ\rho)^*F_\CK}
\cong\,\ee^{-\frac{\ii}{2}\int_{h_0}\Phi^*H}\,\ee^{\ii\sum\limits_{c\subset h_0}
\int_cs_c^*B\,+\,\ii\sum\limits_{\rho(c)\subset h_0}\int_cs_c^*B}\cr\cr
&&\times\,\Big(\mathop{\otimes}\limits_{b\subset c\subset h_0}
hol_\CL(s_c|_b,s_{h_0}|_b)\Big)\otimes\Big(\mathop{\otimes}
\limits_{\rho(b)\subset c\subset h_0}hol_\CL(s_c\circ\rho|_b,s_{h_0}\circ\rho|_b)\Big)
\otimes\,\Big(\mathop{\otimes}\limits_{b\subset\rho(c)\subset h_0}
hol_\CK(s_c\circ\rho|_b,s_{h_0}|_b)\Big)\cr\cr
&&\otimes\Big(\mathop{\otimes}
\limits_{\rho(b)\subset\rho(c)\subset h_0}hol_\CK(s_c|_b,s_{h_0}\circ\rho|_b)\Big)
\,\cong\,\ee^{-\frac{\ii}{2}\int_{h_0}\Phi^*H}\,\ee^{\ii\sum\limits_{c\subset h_0}
\int_cs_c^*B\,+\,\ii\sum\limits_{\rho(c)\subset h_0}\int_cs_c^*B}\cr\cr
&&\times\,\Big(\mathop{\otimes}\limits_{b\subset c\subset h_0}
\big(hol_\CL(s_c|_b,s_b)\otimes hol_\CL(s_b,s_{h_0}|_b)\big)\Big)
\otimes\Big(\mathop{\otimes}\limits_{\rho(b)\subset c\subset h_0}
\big(hol_\CK(s_c\circ\rho|_b,s_b)\otimes
hol_\CK(s_b,s_{h_0}\circ\rho|_b)\big)\Big)\cr
&&\otimes\,\Big(\mathop{\otimes}\limits_{b\subset\rho(c)\subset h_0}\hspace{-0.15cm}
\big(hol_\CK(s_c\circ\rho|_b,s_b)\otimes hol_\CL(s_b,s_{h_0}|_b)\big)\Big)
\otimes\Big(\mathop{\otimes}
\limits_{\rho(b)\subset\rho(c)\subset h_0}\hspace{-0.15cm}\big(hol_\CL(s_c|_b,s_b)\otimes
hol_\CK(s_b,s_{h_0}\circ\rho|_b)\big)\Big).\qquad
\qqq
But
\qq
&&\ee^{\ii\sum\limits_{c\subset h'_0}\int_ci_c^*B}
\Big(\mathop{\otimes}\limits_{b\subset c\subset h'_0}hol_\CL(s_c|_b,s_b)\Big)
\otimes\Big(\mathop{\otimes}\limits_{\rho(b)\subset c\subset h'_0}
hol_\CK(s_c\circ\rho|_b,s_b)\Big)\cr
&&=\ee^{-\ii\sum\limits_{\rho(c)\subset h_0}\int_ci_c^*B}
\Big(\mathop{\otimes}\limits_{\rho(b)\subset\rho(c)\subset h_0}hol_\CL(s_c|_b,s_b)^{-1}
\big)\otimes\Big(
\mathop{\otimes}\limits_{b\subset\rho(c)\subset h_0}hol_\CK(s_c\circ\rho|_b,s_b)^{-1}
\Big)
\qqq
Hence
\qq
&&\ee^{-\frac{\ii}{2}\int_{h'_0}\Phi^*H\,+\,\ii\sum\limits_{c\subset{h'_0}}
\int_cs_c^*B}\Big(\mathop{\otimes}\limits_{b\subset c\subset h'_0}
hol_\CL(s_c|_b,s_b)\Big)\otimes\Big(\mathop{\otimes}
\limits_{\rho(b)\subset c\subset h'_0}hol_\CK(s_c\circ\rho|_b,s_b)\Big)\cr\cr
&&\cong\,\ee^{-\frac{\ii}{2}\int_{h_0}\Phi^*H\,+\,\ii\sum\limits_{c\subset h_0}
\int_cs_c^*B}\Big(\mathop{\otimes}\limits_{b\subset c\subset h_0}
hol_\CL(s_c|_b,s_b)\Big)\otimes\Big(\mathop{\otimes}
\limits_{\rho(b)\subset c\subset h_0}hol_\CK(s_c\circ\rho|_b,s_b)\Big)\cr
&&\otimes\Big(\mathop{\otimes}\limits_{b\subset c\subset h_0}hol_\CL(s_b,s_{h_0}|_b)\Big)\otimes\Big(\mathop{\otimes}\limits_{b\subset\rho(c)\subset h_0}
hol_\CL(s_b,s_{h_0}|_b)\Big)\cr
&&\otimes\Big(\mathop{\otimes}\limits_{\rho(b)\subset c\subset h_0}hol_\CK(s_b,
s_{h_0}\circ\rho|_b)\Big)\otimes\Big(\mathop{\otimes}
\limits_{\rho(b)\subset\rho(c)\subset h_0}hol_\CK(s_b,
s_{h_0}\circ\rho|_b)\Big)\cr\cr
&&\cong\,\ee^{-\frac{\ii}{2}\int_{h_0}\Phi^*H\,+\,\ii\sum\limits_{c\subset h_0}
\int_cs_c^*B}\Big(\mathop{\otimes}\limits_{b\subset c\subset h_0}
hol_\CL(s_c|_b,s_b)\Big)\otimes\Big(\mathop{\otimes}
\limits_{\rho(b)\subset c\subset h_0}hol_\CK(s_c\circ\rho|_b,s_b)\Big)
\qqq
because the dropped factors receive similar contributions
canceling pairwise as they correspond to opposite orientations of $\,b$. 
\vskip 0.1cm

2. \,If for the triangle $\,c_0\in\Sigma\,$ shared by the tetrahedra
$\,h_1\,$ and $\,\rho(h_2)\,$ 
we change the map $\,s_{c_0}\,$ to $\,s'_{c_0}$, \,or the triangle $\,c_0\,$ to 
$\,c_0'=\rho(c)\subset h_2\,$ and the map $\,s_{c_0}\,$ to $\,s_{c'_0}$, \,or for the edge
$\,b_0\subset\Sigma\,$ we change $\,s_{b_0}\,$ to $\,s'_{b_0}$, \,or $\,b_0\,$
to $\,b_0'=\rho(b_0)\,$ and $\,s_{b_0}\,$ to $\,s_{b_0'}\,$ then
the equivalence of the new expressions for $\,K_\CG(\Phi)\,$ to the old one
is shown the same way as in the proof of Proposition 3 in
Sec.\,\ref{sec:sqrt_hol}.
\vskip 0.1cm

\begin{figure}[!h]
\vskip -0.1cm
\begin{center}
\leavevmode
 \includegraphics[width=6cm,height=4cm]{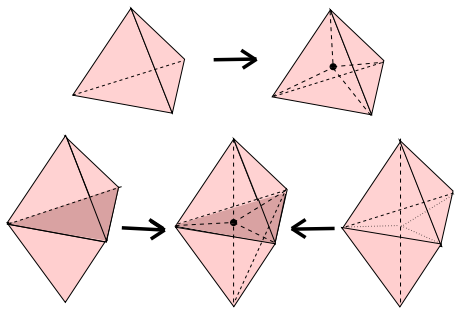}
\vskip -0.1cm
\caption{Three-dimensional Pachner moves}
\label{fig:Pachner3}
\end{center}
\vskip -0.1cm
\end{figure}

3. \,The independence of the trivialization of $\,\CR\,$ lifting
the orbifold triangulation of $\,\tilde\CR\,$ is proven similarly
as in point 3 of the proof of Proposition 3 in Sec.\,\ref{sec:sqrt_hol}
but using now the $\,3d\,$ Pachner moves \cite{Pachner} that subdivide
a tetrahedron into 4 ones or two tetrahedra sharing a triangle
or 3 ones sharing an edge to 6 tetrahedra, see \,Fig.\,\ref{fig:Pachner3},
with the simplest choices of the maps $\,s_h,s_c,s_b\,$ for the new simplices.
We leave the details to the reader.
\vskip -0.1cm

\hspace{14cm}$\blacksquare$

\noindent{\bf Remark.} \,Proposition 6 implies, in particular, that
the right hand side of (\ref{K(Phi)}) does not dependent on the choice
of the submanifold with boundary $\,\CF\subset\CR\,$ forming
the closure of a fundamental domain for the involution $\,\rho\,$
of $\,\CR$. \,Although the right hand side of (\ref{K(Phi)}) may
be often defined using a homotopic non-local formula (\ref{sqrtSWZ})
for the square root of gerbe holonomy, \,the local approach 
based on gerbe theory is useful to establish such a result that,
in application to topological insulators, is a powerful source
of equalities between different forms of invariants
[\onlinecite{GenWar},\,\onlinecite{MT}].

\nsection{Basic gerbe on the group $\,U(N)\,$ and
the time reversal}
\label{sec:basic_gerbe}

\noindent We would like to apply the constructions of the
previous sections to the case when $\,M\,$ is the unitary group
$\,U(N)\,$ in $\,N\,$ dimensions and $\,\CG\,$ is the so called basic
gerbe on $\,U(N)\,$ with curvature given by the closed bi-invariant
3-form 
\qq
H\,=\,\frac{_1}{^{12\pi}}\,\tr(u^{-1}du)^3\,.
\qqq
Different construction of such a gerbe are possible but they all
lead to the same holonomy $\,\Hol_\CQ(\phi)\,$ that is completely fixed 
by Witten's rule (\ref{SWZ}). In the application considered in the sequel, 
we shall, however, also need $\,\sqrt{\Hol_\CG(\phi)}\,$ for 
maps $\,\phi\,$ satisfying the equivariance condition 
(\ref{equivcond}), where $\,\vartheta\,$ is an orientation-preserving
involution of $\,\Sigma\,$ and $\,\Theta\,$ is the involution
on $\,U(N)\,$ generated by the adjoint action 
$\,u\mapsto\theta u\theta^{-1}\,$ of the time reversal
$\,\theta:\mathbb C^N\rightarrow\mathbb C^N$. The transformation
$\,\theta\,$ is an anti-unitary map that squares to $\,\pm I$.
\,There is a problem in applying the construction of
$\,\sqrt{\Hol_\CG(\phi)}\,$ from \,Sec.\,\ref{sec:sqrt_hol}\, for such
an involution $\,\Theta$. \,On the
one hand, when $\,\theta^2=I\,$ then there exists a 
$\,\mathbb Z^2$-equivariant extension $\,\tilde\CG\,$ of the basic
gerbe $\,\CG$. However, in that case the fixed-point set 
$\,U(N)'\subset U(N)\,$ of $\,\Theta\,$ is conjugate to 
the subgroup $\,O(N)\subset U(N)\,$ and is neither connected nor 
simply connected, so that the construction of \,Sec.\,\ref{sec:sqrt_hol}
\,does not work. On the other hand, although for $\,\theta^2=-I\,$
(requiring an even $\,N$) \,the fixed point set $\,U(N)'\,$ 
is conjugate to the subgroup $\,Sp(N)\subset U(N)\,$ and is 1-connected, 
\,there is no $\,\mathbb Z_2$-equivariant extension $\,\tilde\CG\,$ of 
the basic gerbe $\,\CG\,$ in that case. \,This was discussed in
detail in \,Sec.\,I\, of \cite{GenWar} using the equivalent concept of 
a $\,\Zb$-equivariant structure on the gerbe $\,\CG$.
\,In \cite{GenWar} a slight modification of the construction of $\,\CG\,$ 
from \cite{MS} was used, \,see \,Sec.\,H \,of \cite{GenWar}, but 
the conclusions are independent of the choice of the basic gerbe
$\,\CG\,$ on $\,U(N)$. 
\vskip 0.1cm

What partially saved the day in the case when $\,\theta^2=-I\,$ 
was the passage to the double-cover group
\qq
\hat U(N)\,=\,\big\{(u,\omega)\in U(N)\times U(1)\,\big|\,\omega^2
=\det(u)\big\},
\qqq
with the lift $\,\hat\Theta(u,\omega)=(\Theta(u),\omega^{-1})\,$
of the involution $\,\Theta$. The covering map from $\,\hat U(N)\,$ 
to $\,U(N)\,$ just forgets $\,\omega\,$ so that the corresponding deck 
transformation of $\,\hat U(N)\,$ is given by the multiplication by 
$\,(I,-1)$. \,Let $\,\hat\CG\,$ be the pullback to $\,\hat U(N)\,$ 
by the covering map of the basic gerbe $\,\CG\,$ on $\,U(N)\,$ 
(obtained by naturally pulling back all the elements of the structure 
of $\,\CG$)\footnote{The gerbe $\hat\CG\,$ introduced here should not 
be confused with the quotient gerbes discussed at the end of 
\,Sec.\,\ref{sec:equiv_gerbes} \,for which
we used the same notation.}. 
If $\,\hat U(N)\,$ is considered with the $\,\Zb$-action
induced by $\,\hat\Theta\,$ then, as discussed in \,Sec.\,J \,of 
\cite{GenWar}, 
\,there exists a $\,\Zb$-equivariant structure on $\,\hat\CG$, \,or, 
equivalently - see Appendix - a $\,\Zb$-equivariant extension 
$\,\tilde{\hat\CG}\,$ of the gerbe $\,\hat\CG$. 
\,Besides, if the involution $\,\vartheta\,$ on $\,\Sigma\,$ corresponds
to a hyperelliptic cover, \,see \,Sec.\,\ref{sec:surf_invol}, \,then any map 
$\,\phi:\Sigma\rightarrow U(N)\,$ satisfying the equivariance 
condition (\ref{equivcond}) has $\,\det(\phi)\,$ that winds even number of
times along the cycles of $\,\Sigma$. \,It follows that $\,\phi\,$ lifts
to a map $\,\hat\phi:\Sigma\rightarrow\hat U(N)$. Besides, the lift
$\,\hat\phi\,$ satisfies the equivariance condition relative to the involution 
$\,\hat\Theta\,$ and is unique up to a composition with 
the multiplication by $\,(I,-1)\,$ in $\,\hat U(N)$. 
\,The fixed-point set $\,\hat U(N)'\subset \hat U(N)\,$ of $\,\hat\Theta\,$
is composed of two disjoint simply connected components, one isomorphic 
to $\,Sp(N)\,$ and the other obtained by the deck transformation of 
the first one. Although $\,\hat U(N)'\,$ is not 1-connected and the 
trivialization of the flat line bundle $\,N'\,$ over $\,\hat U(N)'\,$ 
used in the definition of the square root of gerbe holonomy in 
\,Sec.\,\ref{sec:sqrt_hol} \,has independent sign ambiguities on each 
connected component of $\,\hat U(N)'$, \,it was shown in \,Sec.\,F \,of
\cite{GenWar} that such ambiguities cancel for the maps 
$\,\hat\phi\,$ as above allowing an unambiguous definition 
of $\,\sqrt{\Hol_{\hat\CG}(\hat\phi)}$. \,Furthermore, the latter quantity 
is equal for the two possible choices of the lift $\,\hat\phi$. It was used 
in \cite{GenWar} as a definition of $\,\sqrt{\Hol_\CG(\phi)}\,$ for the maps 
$\,\phi:\Sigma\rightarrow U(N)\,$ satisfying the equivariance condition 
(\ref{equivcond}) if $\,\vartheta\,$ corresponds to a hyperelliptic cover 
and $\,\Theta\,$ is given by the adjoint action of the time 
reversal $\,\theta\,$ squaring to $\,-I$. \,In particular, this covers
the case of the 2-torus $\,\mathbb T^2\,$ with the involution
$\,\vartheta\,$ induced by $\,k\mapsto-k$, \,see
\,Sec.\,\ref{sec:surf_invol} \,above. It follows by a simple extension
of the results of \cite{CDFGT} that any map $\,\phi:\mathbb T^2
\rightarrow U(N)\,$ equivariant with respect to the $\,k\mapsto-k\,$
and the time-reversal involutions, the latter with $\,\theta^2=-I\,$ can be
extended after a composition with an $\,SL(2,\mathbb Z)\,$ diffeomorphism
of $\,\mathbb T^2\,$ (in order to render the winding of $\,\det(\phi)\,$
around the $\,k_1$-cycle trivial) to an equivariant map
$\,\psi:\CT=D\times S\rightarrow U(N)$, \,where $\,\CT\,$ is taken
with the involution $\,(z,v)\mapsto(\bar z,\bar v)$, \,see the beginning of
\,Sec.\,\ref{sec:homot_form}. Again, $\,\psi\,$ may be lifted to an
equivariant map $\,\hat\psi:\CT\rightarrow\hat U(N)$. Then the application
of (\ref{sqrtD}) and of Proposition 4 shows that the equality
(\ref{sqrtSWZ}) holds in that case.
\vskip 0.1cm

A similar construction permits to define unambiguously the $\,3d$-index
$\,K_\CG(\Phi)\,$ of \,Sec.\,\ref{sec:3d_index} \,for the maps 
$\,\Phi:\mathbb T^3\rightarrow U(N)\,$ satisfying the equivariance condition 
(\ref{equivPhi}) if the involution $\,\rho\,$ of the 3-torus 
$\,\mathbb T^3\,$ is induced by $\,k\mapsto-k\,$ and $\,\Theta\,$
is given by the adjoint action of the time reversal $\,\theta\,$ with 
$\,\theta^2=-I$, \,see again Sec. F of \cite{GenWar}. \,One shows that 
$\,\Phi\,$ lifts to the map $\,\hat\Phi:\mathbb T^3\rightarrow\hat U(N)\,$ 
equivariant with respect to $\,\hat\Theta\,$ and that 
$\,K_{\hat\CG}(\hat\Phi)\,$ is unambiguously defined and independent of 
the choice of the lift $\,\hat\Phi$.

\nsection{Applications to the time-reversal invariant topological insulators}
\label{sec:appli_TI}

\noindent The $2^{\rm nd}$ part of lectures \cite{GenWar} discussed how
the square root of the gerbe holonomy and the corresponding $\,3d$-index
provide invariants of the topological time-reversal-symmetric insulators
in two and three space dimensions. For completeness, we shall list those
results here.
\vskip 0.1cm

In the simplest case, the $d$-dimensional insulators are described 
by Hamiltonians on a crystalline lattice that, after the discrete Fourier-Bloch 
transformation, give rise to a (smooth) map
\qq
\mathbb T^d\ni k\,\longmapsto\,h(k)=h(k)^\dagger\,\in\,End(\mathbb C^N)
\qqq
and all the hermitian matrices $\,h(k)\,$ have a spectral gap around 
the Fermi energy $\,\epsilon_F$. \,Denote
by $\,p(k)\,$ the spectral projector on the eigenstates of $\,\,h(k)\,$ 
with energies $\,<\epsilon_F\,$ which then depends smoothly on
$\,k\in\mathbb T^d$.
\vskip 0.1cm

For the electronic time-reversal-symmetric insulators,
\qq
\theta h(k)\theta^{-1}=h(-k)\qquad{\rm and}\qquad\theta p(k)\theta^{-1}=p(-k)\,,
\qqq
where $\,\theta:\mathbb C^N\rightarrow\mathbb C^N\,$ is an anti-unitary
with $\,\theta^2=-I$. \,Denote by $\,u_p(k)\,$ the unitary matrix 
$\,I-2p(k)$. \,In two or three dimensions, the map
$\,\mathbb T^d\ni k\ \longmapsto\ u_p(k)\in U(N)\,$ is then 
equivariant, \,i.e. $\,\Theta(u_p(k))=u_p(-k)$, \,where 
$\,\Theta(u)=\theta u\theta^{-1}$. \,This is still the case 
for the restriction  of the three dimensional $\,u_p\,$ to
any 2-torus $\,\mathbb T^2\subset\mathbb T^3\,$ preserved by the
$\,k\mapsto-k\,$ involution of $\,\mathbb T^3$. 
\vskip 0.4cm

\noindent{\bf Proposition 7} [\onlinecite{CDFG},\,\onlinecite{CDFGT},\,\onlinecite{GenWar}]{\bf.}
\,Let $\,\CG\,$ be the basic gerbe on $\,U(N)$.

\noindent 1. \,For $\,d=2$, $\,\sqrt{Hol_\CG(u_p)}=(-1)^{\KM}$, \,where 
$\,\KM\in\{0,1\}\,$ is the Fu-Kane-Mele invariant
[\onlinecite{KaneMele},\,\onlinecite{FK}] 
of the time-reversal-symmetric $2d$ topological insulators.

\noindent 2. \hspace{0.14cm}For $\,d=3$, $\,K_\CG(u_p)=(-1)^{\KM^s}\ $ where 
$\,\KM^s\in\{0,1\}\,$ is the ``strong'' Fu-Kane-Mele invariant 
\cite{FKM} of the time-reversal-symmetric $3d$ topological insulators.
\vskip 0.4cm

\begin{figure}[h]
\vskip -0.1cm
\begin{center}
\leavevmode
 \includegraphics[width=5.7cm,height=4.5cm]{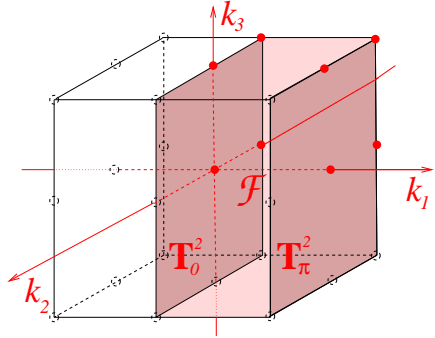}
\vskip -0.1cm
\caption{$\mathbb T^3\,$ viewed as the periodized cube with the 
domain $\,\CF\,$ bounded by 2-tori $\,\mathbb T^2_0\,$ and 
$\,\mathbb  T^2_\pi$}
\label{fig:torus3d}
\end{center}
\end{figure}

\noindent{\bf Remark.} \,The expression (\ref{K(Phi)}) for $\,K_\CG(u_p)\,$
for $\,\CF\subset\mathbb T^3\,$ bounded by the 2-tori $\,\mathbb T^2_0\,$
and $\,\mathbb T^2_\pi$, \,see Fig.\,\ref{fig:torus3d}, \,leads to the 
relation $\,\KM^s=\KM|_{\mathbb T^2_0}+\KM|_{\mathbb T^2_\pi}$, \,known from 
\cite{FKM}, \,between the strong and the weak Fu-Kane-Mele invariants, the
latter defined for the involution-preserved $\,\mathbb T^2\subset
\mathbb T^3\,$ by the relation $\,(-1)^{\KM|_{\mathbb T^2}}
=\sqrt{\Hol_\CG(u_p|_{\mathbb T^2})}$. \,Of course, the pair $\,(\mathbb T^2_0,
\,\mathbb T^2_\pi)\,$ could be replaced by similar pairs orthogonal to
other axes.

Further applications concern the so called Floquet systems described 
by lattice Hamiltonians periodically depending on time. After the 
discrete Fourier-Bloch transformation, such a Hamiltonian gives rise 
to a map
\qq
\mathbb R\times\mathbb T^d\ni(t,k)\,\longmapsto\,h(t,k)=h(t,k)^\dagger
=h(t+2\pi,k)\,\in\,End(\mathbb C^N)\,,
\qqq
where, for convenience, we fixed the period of temporal driving to $\,2\pi$.
\,The time evolution of the corresponding systems is described by the
unitary matrices $\,u(t,k)\,$ such that
\qq
\ii\partial_tu(t,k)=h(t,k)\,u(t,k),\qquad u(0,k)=I,\qquad u(t+2\pi,k)
=u(t,k)\,u(2\pi,k)\,.
\qqq 
The Floquet theory that deals with such systems is based on the
diagonalization of the unitary matrices $\,u(2\pi,k)\,$ whose eigenvalues
are written as $\,\ee^{-\ii e_n(k)}$, \,where $\,e_n(k)\,$ are called the (band)
``quasienergies''. Let us suppose that $\,\epsilon\in[-2\pi,0[\,$ is such that
$\,\ee^{-\ii\epsilon}\,$ is not in the spectrum of $\,u(2\pi,k)\,$ for all $\,k\,$
(i.e. $\epsilon\,$ is in the quasienergy gap). 
\,Then the ``effective Hamiltonian''
\qq
h_\epsilon(k)\equiv h_\epsilon(u(2\pi,k))=\frac{_\ii}{^{2\pi}}\,
\ln_{-\epsilon}(u(2\pi,k))\,,
\qqq
where, by definition, $\,\ln_{-\epsilon}(\ee^{\ii\varphi})=\ii\varphi\,$ 
if $\,-\epsilon-2\pi<\varphi<-\epsilon$, \,is well defined and depends
smoothly on $\,k\in\mathbb T^d$. \,It satisfies
$\,u(2\pi,k)=\ee^{-2\pi\ii\,h_\epsilon(k)}.$
\,For two gap quasienergies $\,-2\pi\leq\epsilon\leq\epsilon'<0$,
\qq
h_{\epsilon'}(k)-h_{\epsilon}(k)
=p_{\epsilon,\epsilon'}(u(2\pi,k))\equiv p_{\epsilon,\epsilon'}(k)\,,
\qqq
where $\,p_{\epsilon,\epsilon'}(k)\,$ is the spectral projector of $\,u(2\pi,k)\,$
on quasienergies $\,\epsilon<e_n(k)<\epsilon'$.
One may use the effective Hamiltonians $\,h_\epsilon(k)\,$ to introduce
the periodized evolution operators
\qq
v_\epsilon(t,k)=u(t,k)\,\ee^{\ii t h_\epsilon(k)}=v_\epsilon(t+2\pi,k)
\qqq
that define a map $\,v_\epsilon:\mathbb T^{d+1}\rightarrow U(N)$. 
\vskip 0.1cm

For the electronic time-reversal-symmetric Floquet systems,
\qq
\theta h(t,k)\theta^{-1}=h(-t,-k)
\qqq
for an anti-unitary $\,\theta\,$ with $\,\theta^2=-I$. \,It follows
then that
\qq
\Theta(v_\epsilon(t,k))\equiv\theta v_\epsilon(t,k)\theta^{-1}
=v_\epsilon(-t,-k)\qquad{\rm and}\qquad
\theta p_{\epsilon,\epsilon'}(k)\theta^{-1}=p_{\epsilon,\epsilon'}(-k)\,.
\qqq
In particular, \,for $\,-2\pi\leq\epsilon\leq\epsilon'<0$,
\,one may consider the Fu-Kane-Mele invariants $\,\KM_{\epsilon,\epsilon'}\,$
and $\,\KM^s_{\epsilon,\epsilon'}\,$ of the quasienergy bands between 
$\,\epsilon\,$ and $\,\epsilon$, \,defined in two
and three dimensions, \,respectively, \,by the relations
\qq
(-1)^{\KM_{\epsilon,\epsilon'}}=\sqrt{\Hol_{\CG}(u_{p_{\epsilon,\epsilon'}})}\,,
\qquad(-1)^{\KM^s_{\epsilon,\epsilon'}}=K_\CG(u_{p_{\epsilon,\epsilon'}})\,,
\qqq
where, as before, $\,\CG\,$ is the basic gerbe on $\,U(N)\,$ and 
$\,u_{p_{\epsilon,\epsilon'}}(k)=I-2p_{\epsilon,\epsilon'}(k)$.
\vskip 0.1cm

In [\onlinecite{CDFG},\,\onlinecite{CDFGT}] and \cite{GenWar} additional
dynamical invariants $\,K_\epsilon\,$ and $\,K^s_\epsilon\,$ with values in\
$\,\{0,1\}\,$ were introduced for the time-reversal-invariant 
Floquet systems in two and three dimensions, \,respectively, \,such that 
\qq
(-1)^{K_\epsilon}=K_\CG(v_\epsilon)\,,\qquad (-1)^{K^s_\epsilon}
=K_\CG(v_\epsilon|_{t=\pi})\,.
\qqq
In $\,3d$, \,one can also define weak dynamical invariants 
$\,K_\epsilon|_{\mathbb T^2}\,$ for 2-tori $\,\mathbb T^2\subset T^3\,$ preserved
by the $\,k\mapsto-k\,$ involution of $\,\mathbb T^3\,$ setting
\qq
(-1)^{K_\epsilon|_{\mathbb T^2}}=K_\CG(v_\epsilon|_{(\mathbb R/2\pi\mathbb Z)
\times\mathbb T^2})\,.
\qqq 
One has the following
\vskip 0.4cm

\noindent{\bf Proposition 8} [\onlinecite{CDFG},\,\onlinecite{CDFGT},\,\onlinecite{GenWar}]{\bf.} 

\noindent 1. \,(Relation between strong and weak invariants).
\qq
K^s_\epsilon=K_\epsilon|_{\mathbb T^2_\pi}-K_\epsilon|_{\mathbb T^2_0}\ \,{\rm mod}\ 2\,.
\qqq

\noindent 2. \hspace{0.08cm}(Relation to the Fu-Kane-Mele invariants).
\,For two gap quasienergies $\,-2\pi\leq\epsilon\leq\epsilon'<0$,
\qq
K_{\epsilon'}-K_\epsilon=\KM_{\epsilon,\epsilon'}\,,\qquad 
K^s_{\epsilon'}-K^s_\epsilon=\KM^s_{\epsilon,\epsilon'}\,.
\qqq
\vskip 0.4cm

\noindent {\bf Remark.} \,The invariants $\,K_\epsilon\,$ are the counterparts 
for time-reversal-symmetric gapped Floquet systems of
the dynamical invariants for such systems without
time-reversal symmetry introduced in \cite{RLBL}, see also \cite{NR}.
They are supposed to count modulo 2 the ``Kramers pairs'' (related by
the time-reversal) of eigenstates of the evolution operator
over one period on a half-lattice that have quasienergy $\,\epsilon\,$
and are localized near the lattice edge \cite{CDFG}.
\vskip 0.5cm

\setcounter{section}{0}
\setcounter{equation}{0} 
\def\theequation{A.\arabic{equation}}
\centerline{\bf\small APPENDIX}
\vskip 0.5cm

\noindent We shall describe here the relation between a
$\,\mathbb Z_2$-extension $\,\tilde\CG=(\tilde Y,\tilde B,\tilde\CL,\tilde t)\,$
of the gerbe $\,\CG=(Y,B,\CL,t)\,$ and a $\,\mathbb Z_2$-equivariant
structure on $\,\CG\,$ defined in \cite{GenWar} following
\cite{GSW}. The definition given in \cite{GenWar} presupposed a simplified
situation when the involution $\,\Theta$, \,inducing the $\,\mathbb Z_2$-action
on the base space $\,M$, \,lifts to an involutive map
$\,\Theta_Y\,$ of $\,Y\,$ (that induces involutions
$\,\Theta_Y^{[p]}\,$ on $\,Y^{[p]}$). The $\,\mathbb Z_2$-equivariant
structure on $\,\CG\,$ was specified in \cite{GenWar} \,by a line-bundle 
$\,\CN\,$ over $\,Y\,$ with curvature $\,\Theta_Y^*B-B\,$ and an
isomorphism $\,\nu:\CL\otimes p_2^*\CN\rightarrow
p_1^*\CN\otimes(\Theta_Y^{[2]})^*\CL\,$ of line-bundles over $\,Y^{[2]}\,$
that commutes with the groupoid multiplication $\,t\,$ in $\,\CL$.
\,Furthermore, \,the flat line bundle $\,\CN\otimes\Theta_Y^*\CN\equiv\CQ\,$
was assumed to be equipped with a trivializing section $\,S\,$ such that
$\,\Theta_\CQ\circ S=S\circ\Theta\,$ for the involutive isomorphism
$\,\Theta_\CQ\,$ of the line bundle $\,\CQ\,$ defined by the permutation
of the tensor factors that lifts the involution
$\,\Theta_Y$ on the base, \,see Sec.\,C and E of \cite{GenWar}. From those
data, one may recover the line bundle $\,\tilde\CL\,$ of gerbe $\,\tilde\CG\,$
setting 
\qq
&&\tilde\CL_{(1,y_1),(1,y_2)}=\CL_{y_1,y_2}=\tilde\CL_{(-1,y_1),(-1,y_2)}\,,\\
&&\tilde\CL_{(1,y_1),(-1,y_3)}=\CN_{y_1}\otimes\CL_{\Theta_Y y_1,y_3}
=\tilde\CL_{(-1,y_1),(1,y_3)}\label{A.2}
\qqq
for $\,\pi(y_1)=\pi(y_2)=\Theta(\pi(y_3))$,
\,with the obvious $\,\mathbb Z_2$-symmetry.
\,The groupoid multiplication $\,\tilde t\,$ is then defined by
the linear maps
\qq
&&\tilde\CL_{(1,y_1),(1,y_2)}\otimes\tilde\CL_{(1,y_2),(1,y_2')}=
\CL_{y_1,y_2}\otimes\CL_{y_3,y_2'}\,\mathop{\longrightarrow}\limits^{t}\,
\CL_{y_1,y_2'}=\tilde\CL_{(1,y_1),(1,y_2')}\,,\\ 
&&\tilde\CL_{(1,y_1),(1,y_2)}\otimes\tilde\CL_{(1,y_2),(-1,y_3)}=
\CL_{y_1,y_2}\otimes\CN_{y_2}\otimes\CL_{\Theta_Yy_2,y_3}\,
\mathop{\longrightarrow}\limits^{\nu\otimes\Id}\,\CN_{y_1}\otimes
\CL_{\Theta_Yy_1,\Theta_Yy_2}\otimes\CL_{\Theta_Yy_2,y_3}\cr
&&\hspace{8.3cm}\mathop{\longrightarrow}
\limits^{\Id\otimes t}\,\CN_{y_1}\otimes\CL_{\Theta_Yy_1,y_3}
=\tilde\CL_{(1,y_1),(-1,y_3)}\,,\\
&&\tilde\CL_{(1,y_1),(-1,y_3)}\otimes\tilde\CL_{(-1,y_3),(-1,y'_3)}=
\CN_{y_1}\otimes\CL_{\Theta_Yy_1,y_3}\otimes\CL_{y_3,y'_3}\,\mathop{\longrightarrow}
\limits^{\Id\otimes t}\,\CN_{y_1}\otimes\CL_{\Theta_Yy_1,y'_3}
=\tilde\CL_{(1,y_1),(-1,y'_3)}\,,\\
&&\tilde\CL_{(1,y_1),(-1,y_3)}\otimes\tilde\CL_{(-1,y_3),(1,y'_1)}=
\CN_{y_1}\otimes\CL_{\Theta_Yy_1,y_3}\otimes\CN_{y_3}\otimes\CL_{\Theta_Yy_3,y'_1}\cr
&&\hspace{3.7cm}\mathop{\longrightarrow}\limits^{\Id\otimes\nu\otimes\Id}\,
\CN_{y_1}\otimes\CN_{\Theta_Yy_1}\otimes\CL_{y_1,\Theta_Yy_3}\otimes
\CL_{\Theta_Yy_3,y'_1}\,\mathop{\longrightarrow}\limits^{S\otimes t}\,\CL_{y_1,y'_1}
=\tilde\CL_{(1,y_1),(1,y'_1)}
\qqq
and the $\,\mathbb Z_2$-symmetry of $\,\tilde\CL$. 
\vskip 0.1cm

Conversely, \,given the $\,\mathbb Z_2$-equivariant extension $\,\tilde\CG\,$
of gerbe $\,\CG$, \,we may obtain a $\,\mathbb Z_2$-equivariant structure
on $\,\CG\,$ by setting $\,\CN_y=\tilde\CL_{(1,y)(-1,\Theta_Yy)}\,$ and defining
the line-bundle isomorphism $\,\nu\,$ by the linear maps on the fibers
\qq
&&\CL_{y_1,y_2}\otimes\CN_{y_2}=\tilde\CL_{(1,y_1),(1,y_2)}\otimes
\tilde\CL_{(1,y_2),(-1,\Theta_Yy_2)}\,\mathop{\longrightarrow}\limits^{\tilde t}
\tilde\CL_{(1,y_1),(-1,\Theta_Yy_2)}\cr
&&\mathop{\longrightarrow}\limits^{\tilde t^{-1}}\,
\tilde\CL_{(1,y_1)(-1,\Theta_Yy_1)}\otimes\tilde\CL_{(-1,\Theta_Yy_1),(-1,\Theta_Yy_2)}
\,\,\mathop{\longrightarrow}\limits^{\Id\otimes(-1)\cdot}\CN_{y_1}\otimes
\CL_{\Theta_Yy_1,\Theta_Yy_2}\,.
\qqq
The trivialization of the line bundle $\,\CQ=\CN\otimes\Theta_Y^*\CN\,$
defined by its section $\,S\,$ is then given by
\qq
&&\qquad\CN_y\otimes\CN_{\Theta_Yy}=\tilde\CL_{(1,y),(-1,\Theta_Yy)}\otimes
\tilde\CL_{(1,\Theta_Yy),(-1,y)}\cr
&&\mathop{\longrightarrow}\limits^{\Id\otimes(-1)\cdot}
\,\tilde\CL_{(1,y),(-1,\Theta_Yy)}\otimes
\tilde\CL_{(-1,\Theta_Yy),(1,y)}\,\mathop{\longrightarrow}\limits^{\tilde t}
\tilde\CL_{(1,y),(1,y)}\,\cong\,\mathbb C\,.
\qqq
\vskip 0.1cm

If, \,as in Sec.\,\ref{sec:loc_data}, $\,(O_i)_{i\in J}\,$ is a sufficiently
fine $\,\Theta$-invariant open covering of $\,M\,$ and the maps
$\,s_i:O_i\rightarrow Y\,$ such that $\,\pi\circ s_i=\Id|_{O_i}\,$ obey
the relation $\,s_{-i}=\Theta_Y\circ s_i\circ\Theta\,$ then 
the collection $\,(\Pi_i,\chi_{i_1i_2},f_i)\,$ with the entries
defined in \,Sec.\,\ref{sec:loc_data} \,provides local data
for the $\,\mathbb Z_2$-equivariant structure on the gerbe $\,\CG\,$
\cite{GSW}. In particular, \,the maps $\,\tilde\sigma_{i\,i}^{1(-1)}\,$ introduced
there take values in the line bundle $\,\CN\,$ and
\qq
&\ii\nabla_{\CN}\tilde\sigma_{ii}^{1(-1)}=\Pi_i\,\sigma_{ii}^{1(-1)},\qquad
\nu\big(\sigma_{i_1i_2}\otimes\tilde\sigma_{i_2i_2}^{1(-1)}\big)=\chi_{i_1i_2}\,
\sigma_{i_1i_1}^{1(-1)}\otimes\big(\sigma_{(-i_1)(-i_1)}\circ\Theta\big)\,,&\\
&\tilde\sigma^{1(-1)}_{i\,i}\otimes\big(\tilde\sigma^{1(-1)}_{(-i)(-i)}\circ\Theta\big)
=f_i\,S\,.&
\qqq
\vskip 0.1cm

The definition of $\,\sqrt{\Hol_\CG(\phi)}\,$ in Sec.\,F of \cite{GenWar},
based on the use of $\,\mathbb Z_2$-equivariant structure on $\,\CG$,
\,is equivalent to the one from \,Sec.\,\ref{sec:sqrt_hol}
\,of the present paper for the lifts $\,c\,$ of triangles $\,\tilde c\,$ 
forming the domain $\,F\,$ and lifts $\,b\,$ of edges $\,\tilde b\,$
either shared by two triangles of $\,F\,$ or belonging to the curves 
$\,\ell\,$ such that $\,\partial F=\ell\cup\vartheta(\ell)\,$
and $\,\partial\ell\subset\Sigma'$. \,Indeed, \,for such a choice
of $\,c\,$ and $\,b\,$ the expression (\ref{expression}) reduces to
\qq
&&\quad\ee^{\ii\sum\limits_{c\subset F}\int_cs_c^*B}\Big(\mathop{\otimes}
\limits_{\substack{b\subset c\subset F\\ \vartheta(b)\not\subset\ell}}hol_\CL(s_c|_b,s_b)\Big)
\otimes\Big(\mathop{\otimes}\limits_{b\subset\ell\subset\partial F}hol_{\CK}(s_b,s_c
\circ\vartheta|_b)\Big)\cr
&&\cong\,
\ee^{\ii\sum\limits_{c\subset F}\int_cs_c^*B}\Big(\mathop{\otimes}
\limits_{\substack{b\subset c\subset F\\ \vartheta(b)\not\subset\ell}}hol_\CL(s_c|_b,s_b)\Big)
\otimes\Big(\mathop{\otimes}\limits_{b\subset\ell\subset\partial F}
\big(hol_\CN(s_b)\otimes
hol_\CL(\Theta_Y\circ s_b,s_c\circ\vartheta|_b)\big)\Big)\cr
&&\cong\,
\ee^{\ii\sum\limits_{c\subset F}\int_cs_c^*B}\Big(\mathop{\otimes}
\limits_{\substack{b\subset c\subset F\\ \vartheta(b)\not\subset\ell}}hol_\CL(s_c|_b,s_b)\Big)
\otimes\Big(\mathop{\otimes}\limits_{b\subset\ell\subset\partial F}
\big(hol_\CN(s_b)\otimes
hol_\CL(s_{\vartheta(b)},s_c|_{\vartheta(b)})\big)\Big)\cr
&&=\,\ee^{\ii\sum\limits_{c\subset F}\int_cs_c^*B}\Big(\mathop{\otimes}
\limits_{b\subset c\subset F}hol_\CL(s_c|_b,s_b)\Big)
\otimes\Big(\mathop{\otimes}\limits_{b\subset\ell\subset\partial F}
hol_\CN(s_b)\Big),
\label{A.11}
\qqq
where $\,c\,$ and $\,b\,$ run here over the triangles and edges
of the triangulation of $\,\Sigma\,$ with specified restrictions
that determine their orientations and we used the relation (\ref{A.2})
to represent the line bundle $\,\CK$. \,It was the right hand side of 
(\ref{A.11}) that was used in Sec.\,F of \cite{GenWar} to define 
$\,\sqrt{\Hol_\CG(\phi)}$, employing a trivializing section of the flat
line bundle $\,N'\,$ over $\,M'\,$ whose definition in \cite{GenWar}
agrees with the one given in Sec.\,\ref{sec:sqrt_hol} here.



\vskip 0.5cm

\begin{thebibliography}{natbib}

\bibitem{Alvarez}
O. Alvarez: {\it Topological  quantization  and  cohomology}.
Commun. Math. Phys. {\bf 100} (1985), 279-309

\bibitem{BenBassat}
O. Ben-Bassat: {\it Equivariant gerbes on complex tori}.
J. Geom. Phys. {\bf 64} (2013), 209-221

\bibitem{Bonahon}
F. Bonahon: {\it Geometric structures on 3-manifolds}. In: Handbook
of Geometric Topology, eds. R. B. Sher and R. J. Daverman, Elsevier
Amsterdam 2002, p.\,114

\bibitem{Brylinski}
J.-L. Brylinski: {\it Loop Spaces, Characteristic Classes and Geometric 
Quantization}. Birkhauser, Boston 1993

\bibitem{CMM}
A. L. Carey, J. Mickelsson, M. Murray: {\it Bundle gerbes applied
to quantum field theory}. Rev. Math. Phys. {\bf12} (2000), 65-90 

\bibitem{CDFG}
D. Carpentier, P. Delplace, M. Fruchart and K. Gawędzki: {\it Topological 
index for periodically driven time-reversal- invariant 2D systems}.
Phys. Rev. Lett. {\bf 114} (2015), 106806

\bibitem{CDFGT}
D. Carpentier, P. Delplace, M. Fruchart, K. Gawędzki and C. Tauber: 
{\it Construction and properties of a topological index for periodically 
driven time-reversal invariant 2D crystals}. Nucl. Phys. B {\bf 896} (2015), 
779-834

\bibitem{Deligne}
P. Deligne: {\it Théorie de Hodge : II}. Publ. Math. de l'IHÉS {\bf 40} 
(1971), 5–57

\bibitem{FK}
L. Fu and C. L. Kane: {\it Time reversal polarization and a ${Z}_{2}$ adiabatic
 spin pump}. Phys. Rev. B {\bf 74} (2006), 195312

\bibitem{FKM}
L. Fu, C. L. Kane and E. J. Mele: {\it Topological insulators in three 
dimensions}. Phys. Rev. Lett. {\bf 98} (2007), 106803

\bibitem{KG}
K. Gaw\c{e}dzki: {\it Topological actions in two-dimensional quantum 
field theory}. In:  Non-Perturbative Quantum Field Theory, eds. 
G. 't Hooft, A. Jaffe, G. Mack, P. Mitter et R. Stora, Plenum Press, 
New York, London 1988, pp. 101-142

\bibitem{GenWar}
K. Gaw\c{e}dzki: {\it Bundle gerbes for topological insulators}.
Preprint arXiv:1512.01028 [math-ph], to appear in Proceedings
of Advanced School on Topological Field Theory, Warsaw, December
7-9, 2015

\bibitem{GR}
K. Gaw\c{e}dzki, N. Reis: {\it WZW branes and gerbes}.
Rev. Math. Phys. {\bf 14} (2002), 1281-1334

\bibitem{GSW}
K. Gaw\c{e}dzki, R. R. Suszek and K. Waldorf: {\it Bundle gerbes for 
orientifold sigma models}. Adv. Theor. Math. Phys. {\bf 15} (2011), 621-688 

\bibitem{GSW1}
K. Gaw\c{e}dzki, R. R. Suszek and K. Waldorf: {\it Global gauge anomalies 
in two-dimensional bosonic sigma models}. Commun. Math. Phys. {\bf302} 
(2011), 513-580 

\bibitem{Giraud}
J. Giraud: {\it Cohomologie Non-Ab\'elienne}. Springer 1971 

\bibitem{Gomi}
K. Gomi: {\it Equivariant smooth Deligne cohomology}. Osaka  J. Math.
{\bf 42} (2005), 309-337

\bibitem{Gomi1}
K. Gomi: {\it Relationship between equivariant gerbes and gerbes over 
the quotient space}. Commun. Contemp. Math. {\bf 7} (2005), 207-226

\bibitem{KaneMele}
C. L. Kane and E. J. Mele: {$\mathbb Z_2$ \it topological order and 
the quantum spin Hall effect}. Phys. Rev. Lett. {\bf 95} (2005) 146802

\bibitem{MT}
D. Monaco and C. Tauber: {\it Gauge-theoretic invariants for topological
insulators: A bridge between Berry, Wess-Zumino, and Fu-Kane-Mele}.
Lett. Math. Phys., online first 

\bibitem{Murray}
M. K. Murray: {\it Bundle gerbes}. J. London Math. Soc. {\bf 54} (1996),
403-416

\bibitem{MRSV}
M. K. Murray, D. M. Roberts, D. Stevenson and R. F. Vozzo: 
{\it Equivariant bundle gerbes}. Preprint arXiv: 1506.07931 [math.DG]

\bibitem{Murray-Stevenson}
M. K. Murray, D. Stevenson: {\it Bundle gerbes: stable isomorphism and 
local theory}. J. Lond. Math. Soc. {\bf 62} (2000), 925-937

\bibitem{MS}
M. K.  Murray  and  D. Stevenson:  {\it The  basic  bundle  gerbe  on  
unitary  groups}. J. Geom. Phys. {\bf 58} (2008), 1571-1590

\bibitem{NR}
F. Nathan, M. S. Rudner: {\it Topological singularities and the general 
classification of Floquet-Bloch systems}. New J. Phys. {\bf 17} (2015),
125014

\bibitem{NS}
T. Nikolaus, C. Schweigert: {\it Equivariance in higher geometry}.
Adv. Math. {\bf226} (2011), 3367-3408 

\bibitem{Pachner}
U. Pachner: {\it P.L. homeomorphic manifolds are equivalent by elementary 
shellings}. European J. Combin. {\bf 12} (1991), 129–145

\bibitem{RLBL}
M. S. Rudner, N. H. Lindner, E. Berg and M. Levin: {\it Anomalous edge states 
and the bulk-edge correspondence for periodically driven two-dimensional 
systems}. Phys. Rev. X {\bf 3} (2013), 031005

\bibitem{SSW}
U. Schreiber, C. Schweigert and K. Waldorf: {\it Unoriented WZW models and 
holonomy of bundle gerbes}. Commun. Math. Phys. {\bf 274} (2007), 31-64

\bibitem{WZ}
J. Wess and B. Zumino: {\it Consequences of anomalous Ward identies}.
Phys. Lett. B {\bf37} (1971), 95-97

\bibitem{WittenCA}
E. Witten: {\it Global aspects of current algebra}
Nucl. Phys. B {\bf 223} (1983), 422-432

\bibitem{WittenNA}
E. Witten, {\it Non-abelian bosonization in two dimensions}. Commun. Math.
Phys. {\bf 92} (1984), 455-472 

\end{thebibliography}
\end{document}